\documentclass[11pt,letterpaper]{scrartcl}
\makeatletter
\DeclareOldFontCommand{\rm}{\normalfont\rmfamily}{\mathrm}
\DeclareOldFontCommand{\sf}{\normalfont\sffamily}{\mathsf}
\DeclareOldFontCommand{\tt}{\normalfont\ttfamily}{\mathtt}
\DeclareOldFontCommand{\bf}{\normalfont\bfseries}{\mathbf}
\DeclareOldFontCommand{\it}{\normalfont\itshape}{\mathit}
\DeclareOldFontCommand{\sl}{\normalfont\slshape}{\@nomath\sl}
\DeclareOldFontCommand{\sc}{\normalfont\scshape}{\@nomath\sc}
\makeatother

\usepackage[utf8]{inputenc}
\usepackage[affil-it]{authblk}
\usepackage[hidelinks]{hyperref}
\usepackage{microtype}
\usepackage{xcolor}
\usepackage{csquotes}
\usepackage{cleveref}
\usepackage{siunitx}
\usepackage[numbers,sort&compress]{natbib}
\usepackage[parfill]{parskip}
\usepackage[top=2.5cm, bottom=2.5cm, left=3.2cm, right=3cm]{geometry}
\usepackage{graphicx}
\usepackage{textcomp}

\usepackage{scalefnt}
\newcommand{\abbrev}{\scalefont{.9}}

\newcommand{\STRIPPER}{\text{\abbrev STRIPPER}}

\newcommand{\NLO}{\text{\abbrev NLO}}
\newcommand{\NNLO}{\text{\abbrev NNLO}}
\newcommand{\NNNLO}{\text{\abbrev N$^3$LO}}

\newcommand{\QCD}{\text{\abbrev QCD}}
\newcommand{\EW}{\text{\abbrev EW}}
\newcommand{\QCDEW}{\text{\abbrev QCD-EW}}
\newcommand{\SM}{\text{\abbrev SM}}

\newcommand{\PDF}{\text{\abbrev PDF}}

\newcommand{\CPU}{\text{\abbrev CPU}}
\newcommand{\GPU}{\text{\abbrev GPU}}
\newcommand{\HPC}{\text{\abbrev HPC}}
\newcommand{\LHC}{\text{\abbrev LHC}}

\newcommand{\HLLHC}{\text{\abbrev HL-LHC}}

\newcommand{\IR}{\text{\abbrev IR}}

\usepackage{fancyhdr}
\usepackage[yyyymmdd,hhmmss]{datetime}
\fancypagestyle{plain}
{
	\fancyhf{} 
	\fancyfoot[C]{\textbf{\thepage}}

}
\title{\hfill{\normalfont\small{MSUHEP-22-016}}
\\[2ex]
Computational challenges for\\ multi-loop collider phenomenology\\
	{\large A Snowmass 2021 white paper}
}
\author[1]{Fernando Febres Cordero}
\author[2]{Andreas von Manteuffel}
\author[3]{Tobias Neumann}

\affil[1]{Physics Department, Florida State University, Tallahassee, FL 32306}
\affil[2]{Department of Physics and Astronomy, Michigan State University\\
	East Lansing, MI 48824}
\affil[3]{Department of Physics, Brookhaven National Laboratory, Upton, New York 11973, USA}

\affil[ ]{\newline ffebres@hep.fsu.edu, vmante@msu.edu, tneumann@bnl.gov}

\date{}

\newcounter{notecount}

\makeatletter
\renewcommand\maketitle{
	\begin{center}
		{\Large\bfseries\@title\par\vspace{0.3em}}
		{\scshape\@author\par\vspace{0.3em} \@date}
	\end{center}
}
\makeatother

\begin{document}
	
	\maketitle
    \begin{abstract}
    Precision measurements at the \LHC{} and future colliders require theory predictions with uncertainties at the percent level for many observables. 
    Theory uncertainties due to the perturbative truncation are particularly relevant and must be reduced to fully exploit the physics potential of collider experiments.
	In recent years the theoretical high energy physics community
	has made tremendous analytical and numerical advances to address this challenge.
    In this white paper, we survey state-of-the-art calculations in perturbative quantum field theory for collider phenomenology with a particular focus on the computational requirements at high perturbative orders.
    We show that these calculations can have specific high-performance-computing (\HPC{}) profiles that should to be taken into account in future \HPC{} resource planning.
	\end{abstract}

	\tableofcontents

\section{Introduction}

    Already today, a number of measurements at the \LHC{} reach uncertainties at the percent level. Standard candle 
	processes like $Z$-boson production allow for even better experimental accuracy, down to the per-mille level for normalized kinematic distributions.
	At future colliders a comparable level of precision is expected for a wider range of observables. For example, for the high-luminosity run at the \LHC{} (\HLLHC{}) even rare processes like Higgs boson production require theoretical control of cross sections at the level of 1\%.
	For many processes current theoretical uncertainties do not match the anticipated
	experimental errors.
	To fully exploit the physics potential of current and future collider experiments,
	in particular to unambiguously identify signals of new physics, it is crucial to improve the precision of theoretical predictions. 
	
	Theory predictions for hadron collider phenomenology are based on collinear factorization.
	In this framework predictions are made in terms of 
	parton distribution functions (\PDF{}s) and partonic hard cross sections, and they are expected to be valid up to corrections which are suppressed in the high energy limit.
	The partonic cross sections are computed in perturbation theory, and within the Standard Model of particle physics
	(SM) this means an expansion in the strong and the electroweak couplings.
	A dominant source of uncertainty originates from higher-order terms in these expansions which is the focus of this white paper.
	Nevertheless, we also note that depending on the observable studied and the 
	kinematic region considered, other sources of uncertainty might dominate, for example due to parametric dependence (on \PDF{}s, couplings, masses), non-perturbative
	effects (like hadronization, multi-parton interactions, etc.), or the appearance of large logarithms which would need to be resummed.
	
	In this white paper, we focus on current calculational methods and computational challenges necessary to reduce perturbative truncation uncertainties of parton-level predictions.
	We performed a survey and asked authors of various state-of-the-art \textit{multi-loop} calculations about their computational resources needs.
	For our survey we received responses that cover 53 scientific publications. This data provides a picture of where current analytic and computational requirements lie, and gives an impression of where the field is moving with respect to resource requirements.
	
	In the following, we first give a brief overview of a related study made during the 2013 Snowmass community planning. In section~\ref{sec:integrals} we 
	highlight state-of-the-art methods for the calculation of Feynman integrals, while in section~\ref{sec:amplitudes} we discuss multi-loop
	scattering amplitudes. In section~\ref{sec:applications} we highlight recent precision cross section calculations. We end with conclusions and an outlook 
	in section~\ref{sec:outlook}, where we point out specific computing needs of our community which should be taken into account in future 
	high performance computing (\HPC{}) resource planning.
	
	\paragraph{State-of-the-art at Snowmass 2013.}
	
	In 2013 a white paper on \enquote{Computing for perturbative \QCD{}}~\cite{Hoche:2013zja}\footnote{see also the Snowmass \QCD{} working group 
	report~\cite{Campbell:2013qaa}} presented a survey of computational requirements of then state-of-the-art hadron collider phenomenology. This included benchmarks for high-multiplicity next-to-leading-order (\NLO{}) \QCD {} calculations with up to 6 final-state particles and also benchmarks of early next-to-next-to-leading-order (\NNLO{}) \QCD{} calculations for $2\to 1$ and $2\to 2$ processes.
	
	Concretely, examples presented included an \NLO{} \QCD{} calculation of $W+5$-jet production~\cite{Bern:2013gka} which required about 600,000 \CPU{} hours with year 2013 hardware. 
	Differential \NNLO{} \QCD{} calculations for $W$/$Z$/$H$ production were available~\cite{Melnikov:2006di,Li:2012wna,Anastasiou:2005qj} and benchmarked. 
	Their evaluation took 50,000 core hours (2013), while total inclusive $t\bar{t}$ production~\cite{Barnreuther:2012wtj,Czakon:2013goa} took about 
	1M core hours (2013). Further processes with \NNLO{} QCD corrections included single jet production~\cite{Gehrmann-DeRidder:2013uxn} (85,000 core hours in 2013) 
	and $H+$jet production~\cite{Boughezal:2013uia} (500,000 core hours in 2013) in the gluon-gluon channel. Translating these core hour numbers into node 
	days (100,000 core hours equal about 500 node days assuming 8-core systems from 2013) makes clear that these calculations were only possible due to the use of \HPC{} systems.
	Nevertheless, apart from these examples, the majority of developments at that time did not rely on larger-scale computing resources.
	
	Compared to state-of-the-art hardware from 2013, current state-of-the-art hardware (2022) has single-threading performance that is better by a factor of four to five. 
	Moreover, due to improvements in multi-core architecture, current single \CPU{} (but multi core) node performance is better by a factor of up to 25-30. 
	For example, this alone means that $2\to1$ \NNLO{} \QCD{} calculations that required a large cluster with hundreds of nodes 10 years ago would now run on a small-size
	cluster with just a few nodes. Furthermore, algorithmic and theory developments have considerably brought down computational	requirements, for example allowing (multi-)boson $2\to1$ and $2\to 2$ production at \NNLO{} \QCD{} to be computed within hours on current desktop machines \cite{Campbell:2019dru}. 
	
	In 2013 a community goal was envisioned to take advantage of new large-scale computing and to benefit from using new hardware like \GPU{}s and Intel Phi 
	many-core systems. Further, questions regarding the increased role of parallel computing, what could be gained by consolidation of resources, and 
	limitations in the software environment were posed. Also questions about public availability of codes, grade of automation, expandability, versatility 
	(cuts, etc.) and user-friendliness were raised. All of these are important aspects to ensure that the efforts to complete these challenging calculations
	have the highest impact in the theoretical and experimental communities. 
	These topics continue to be relevant and we will comment on them from today's perspective in this white paper.

	Enormous progress has been achieved since the last Snowmass exercise in the field of precision high energy phenomenology. The very challenging calculations
	from then have become standard and relatively fast, and calculations that then would have appeared as unfeasible have actually been completed, like for example fiducial-level next-to-next-to-next-to-leading-order (\NNNLO{}) \QCD{} calculations for $2\to 1$ processes and \NNLO{} \QCD{} calculation for $2\to 3$ processes.
	In the rest of this article we highlight analytical and numerical techniques that have been developed to allow this to happen and discuss challenges
	for the coming decade in precision \QCD{} phenomenology.
	
\section{Feynman integrals}\label{sec:integrals}

	Feynman or loop integrals are basic building blocks of perturbative quantum field theory
    (for a recent pedagogical introduction see ref.~\cite{Weinzierl:2022eaz}).
	They contain key information about the structure of scattering amplitudes and their singularities. This motivates detailed studies of the mathematical representations of Feynman integrals to gain a deep understanding of their analytical properties and (singular) behaviour in particular kinematic limits.
	For phenomenological purposes, one ultimately wants to
    numerically evaluate the expressions, such that
    the error is reliable, the result has sufficient precision and
    the computation requires acceptable resources.
    In that context, suitable analytical preparations of the
    integrals (and amplitudes) can be
    regarded as effort during the \textit{development} phase,
    while the actual numerical evaluation
    of the expressions during the Monte Carlo integration
    over phase space represent the \textit{production} phase for
    collider observables.
    Different methods have been developed which put more or less
    emphasis on the analytical preparations.
    Typically, dedicated analytical manipulations are somewhat
    process specific and more difficult to automate, while numerical
    techniques allow more flexibility perhaps at the expense of lower
    efficiency. Below we highlight recent progress
    in the development of techniques for the analytic and numerical
    evaluation of Feynman integrals for collider phenomenology.
    
    The current state-of-the-art in evaluating multi-loop Feynman integrals
    include two-loop integrals for $2\to 2$ processes with internal masses (see e.g.~\cite{Heller:2019gkq,Moriello:2019yhu,Agarwal:2020dye,Bronnum-Hansen:2020mzk,Bronnum-Hansen:2021olh,Bronnum-Hansen:2021pqc}),
    $2\to 3$ processes with up to one massive leg (see e.g.~\cite{Papadopoulos:2015jft,Gehrmann:2018yef,Abreu:2020jxa,Canko:2020ylt,Chicherin:2020oor,Badger:2021nhg,Chicherin:2021dyp,Abreu:2021smk,Papadopoulos:2019iam,Kardos:2022tpo}),
    three-loop integrals for $2\to 2$ processes (see e.g.~\cite{Henn:2020lye,Canko:2021xmn}),
    and four-loop integrals for $2\to 1$ processes (see e.g.~\cite{Henn:2016men,Henn:2019rmi,Henn:2019swt,vonManteuffel:2019gpr,vonManteuffel:2020vjv,Agarwal:2021zft,Lee:2021uqq,Lee:2022nhh}). 

    \paragraph{Analytical solutions.}
    In the method of differential equations~\cite{Kotikov:1990kg,Bern:1993kr,Remiddi:1997ny}, one derives
    a coupled, first-order, linear system of differential equations
    for a set of basis or master integrals, where the derivatives
    are taken with respect to the external kinematic invariants.
    To achieve this, one needs to determine linear relations 
    between Feynman integrals, usually computed through integration-by-parts reductions.
    In simpler cases with 
    few physical scales, one
    can solve the differential equations analytically, where
    a suitable choice of basis integrals is essential for the successful
    integration.
    In particular, the choice of a canonical basis~\cite{Henn:2013pwa}
    can greatly simplify the construction of a solution and has been used
    in many calculations in the recent past.
    The direct integration of parametric representations can also
    be used to obtain solutions.
    In the case of linearly reducible integrals~\cite{Brown:2008um},
    the program \texttt{HyperInt}~\cite{Panzer:2014caa} 
    allows for an automated
    approach to obtain results in terms of multiple polylogarithms,
    see e.g.\ \cite{vonManteuffel:2015gxa,Bonetti:2020hqh} for applications.
    
    Obtaining analytical solutions typically requires one to study
    non-standard special mathematical functions in some detail,
    possibly developing new algorithmic methods for them.
    Examples for such functions are harmonic
    polylogarithms~\cite{Remiddi:1999ew},
    multiple polylogarithms~\cite{Remiddi:1999ew,Goncharov:2001iea}
    and elliptic polylogarithms~\cite{BrownLevin,Bloch:2013tra,Adams:2015gva,Ablinger:2017bjx,Remiddi:2017har,Broedel:2017kkb}.
    If a functional basis is chosen with numerical performance
    in mind, the resulting expressions may allow for precise
    and fast numerical evaluation using universal libraries,
    at the order of a second per phase space point for state-of-the-art two-loop amplitudes.
    Several such universal libraries have been developed
    for multiple polylogarithms, for example the implementation of ref.~\cite{Vollinga:2004sn}
    in \texttt{GiNaC} and the \texttt{HandyG} library~\cite{Naterop:2019xaf}.
    More recently, algorithms and tools for iterated integrals related
    to elliptic Feynman integrals have been developed~\cite{Duhr:2019rrs,Walden:2020odh}.
    A more comprehensive discussion of special functions
    relevant for collider physics can be found in a recent Snowmass white paper \cite{Bourjaily:2022bwx}.
    The above methods usually aim at numerical evaluation through series
    expansions of the special functions.
    Alternatively, mapping Chen-iterated integrals to one-fold integral
    representations were shown to provide good performance through numerical
    quadrature, see for example the \texttt{C++} libraries for $2\to 3$ processes in~\cite{Chicherin:2020oor,Chicherin:2021dyp}.

    \paragraph{Numerical methods.}
    In particular for cases with many masses and complicated branch
    cut structure, an analytical solution suitable for numerical applications
    may be difficult to obtain, in particular if intricate analytical continuations
    are required.
    Numerical methods allow to treat these cases and offer great potential for automation.

    A numerical method which has been studied for a long time
    are integrations of Feynman parametric representations.
    Due to the presence of divergences when taking the limit $\epsilon\to 0$ for the dimensional regularization parameter, one
    cannot simply expand the integrand in $\epsilon$ for arbitrary
    integrals.
    One way to make the poles in $\epsilon$ manifest and to arrive at
    finite integrals for the $\epsilon$ expansion which can be
    evaluated on the computer is the technique known as sector decomposition~\cite{Binoth:2000ps}.
    In the past decade, this method has been fully automated
    for physical kinematics~\cite{Borowka:2015mxa} and the public codes
    \texttt{pySecDec}~\cite{Borowka:2018goh} and \texttt{Fiesta}~\cite{Smirnov:2021rhf}
    can efficiently exploit quasi-Monte Carlo methods as well as the usage of \GPU{}s,
    see e.g.~\cite{Borowka:2016ypz,Jones:2018hbb,Chen:2019fla,Chen:2020gae,Agarwal:2020dye} for applications. 
    
    It has also been observed that one can always reduce Feynman integrals
    in terms of a basis of finite Feynman integrals~\cite{vonManteuffel:2014qoa,vonManteuffel:2015gxa,Agarwal:2020dye}.
    In principle, one can straight-forwardly expand these integrals
    in $\epsilon$ and perform the integration numerically.
    In practice, evaluating such finite integrals with
    established sector decomposition codes provides better
    performance than each of these methods alone~\cite{vonManteuffel:2017myy}.
    Still, obtaining sufficient precision for phenomenological applications can
    require substantial computation time and the usage of cluster resources.
    Other numerical approaches include building on established libraries for one-loop integrals to exploit
    dispersion relations to calculate two-loop integrals with internal masses~\cite{Song:2021vru}, and
    using Mellin-Barnes representations to calculate two-loop integrals with several internal masses~\cite{Dubovyk:2018rlg}, among other.

    A semi-analytic alternative can also be constructed by solving differential equations for master integrals
    with respect to external invariants numerically using generalized power series.
    The latter method gained significant popularity in the last few years, since
    it allows not only for a very generic approach but also for particularly precise and fast
    evaluations~\cite{Pozzorini:2005ff,Aglietti:2007as,Lee:2017qql,Heller:2019gkq,Moriello:2019yhu,Frellesvig:2019byn,Heller:2021gun,Abreu:2020jxa,Abreu:2021smk}.
    Typical evaluation times are of the order of a minute per phase-space point
    for two-loop applications.
    The \texttt{DiffExp} program \cite{Hidding:2020ytt} provides a public implementation based on the method
    of Frobenius.
    One can combine this method with the direct integration techniques discussed above
    to fix boundary values in Euclidean phase space points~\cite{Dubovyk:2022frj}.
    
    Another alternative to obtain boundary conditions goes under the name of the auxiliary mass flow method~\cite{Liu:2022mfb,Liu:2017jxz,Liu:2020kpc,Liu:2021wks}.
    It exploits differential equations with respect to a technical mass parameter introduced in the propagators, and a public implementation of the approach, \texttt{AMFlow}~\cite{Liu:2022chg}, has been presented recently.
    Applications of this method can be found for example in refs.~\cite{Bronnum-Hansen:2020mzk,Bronnum-Hansen:2021olh,Bronnum-Hansen:2021pqc}.
    While in the auxiliary mass flow method the required integration-by-parts reductions are more
    challenging (given the modified propagator structure), it has the advantage that it allows to easily fix 
    boundary values in a generic way.

\section{Scattering amplitudes}\label{sec:amplitudes}
	
	Scattering amplitudes are at the core of all collider phenomenology predictions. They contain
	the dynamical information associated with the models used to describe data. At the energies of current high energy colliders, 
	amplitudes are computed perturbatively in the strong and electroweak couplings.
	Quantitatively reliable predictions at the \LHC{} with uncertainties of $\mathcal{O}($15--20$\%)$ require the inclusion 
	of at least first-order corrections in the strong coupling (\NLO{} \QCD{}), while for precision studies with uncertainties below 
	$\mathcal{O}(7\%)$ it is generally necessary to include second-order or higher corrections in
	the strong coupling (\NNLO{} \QCD{}) as well as first-order corrections in the electroweak coupling (\NLO{} electroweak).
	
	To obtain these types of predictions for the wide variety of processes of interest at the \LHC{}, a myriad of 
	one-loop, two-loop and three-loop amplitudes are required (also tree-level amplitudes, which can now be
	computed for essentially arbitrary particle multiplicities with standard tools). Although general calculational methods are available, the complexity of amplitude calculations grows quickly with
	the number of loops as well as with the number of physical scales such as particle masses and kinematic invariants.
	As a consequence, large efforts are required by the 
	high energy theory community to make sure that amplitudes are available for phenomenological applications.
	
	One-loop amplitude calculations in the \SM{} (as well as in theories beyond the \SM{}) can now be completed
	employing highly automated frameworks. Even for whopping $2\rightarrow 8$ processes with multiple 
	internal/external massive particles they can be obtained with standard computational resources. This has been
	the case thanks to the development of highly automated libraries like \texttt{Helac-NLO}~\cite{Bevilacqua:2011xh}, 
	\texttt{MG5\_aMC@NLO}~\cite{Alwall:2014hca,Frederix:2018nkq}, \texttt{NLOX}~\cite{Figueroa:2021txg}, 
	\texttt{OpenLoops}~\cite{Buccioni:2019sur}, \texttt{Recola}~\cite{Denner:2017wsf}, as well as many 
	other public and private tools.
	
	Two- and higher-loop calculations remain challenging, in particular those that are related to many scales. But in the last five years many new calculations have been completed due to
	several major advances in our understanding of the analytic structure of scattering amplitudes in perturbative
	quantum field theory. These calculations include for example two-loop amplitudes for $2\rightarrow 2$ processes with 4 or more scales~\cite{Borowka:2016ehy,Jones:2018hbb,Bonetti:2020hqh,Heller:2020owb,Bronnum-Hansen:2021olh,Bonciani:2021zzf,Chen:2020gae,Agarwal:2020dye,Becchetti:2021axs}, two-loop amplitudes for massless $2\rightarrow 3$ processes~\cite{Badger:2018enw,Abreu:2018zmy,Abreu:2019odu,Badger:2019djh,Abreu:2020cwb,Chawdhry:2020for,Agarwal:2021grm,Abreu:2021oya,Chawdhry:2021mkw,Agarwal:2021vdh}, two-loop amplitudes for $2\rightarrow 3$ processes with
	one external massive particle~\cite{Badger:2021nhg,Badger:2021ega,Abreu:2021asb,Badger:2022ncb}, three-loop form factors for
	$2\rightarrow 1$ processes with massive particles~\cite{Henn:2016tyf,Lee:2018nxa,Ablinger:2018yae,Blumlein:2019oas,Czakon:2021yub,Fael:2022rgm}, three-loop 
	amplitudes for massless $2\rightarrow 2$ processes~\cite{Caola:2020dfu,Caola:2021rqz,Bargiela:2021wuy,Caola:2021izf}, and
	four-loop form factors for $2\rightarrow 1$ processes~\cite{Henn:2016men,Lee:2022nhh,Chakraborty:2022yan}.
	Major progress has been achieved also for the calculation of related quantities, for example the complete five-loop beta function~\cite{Herzog:2017ohr,Luthe:2017ttg} and first results for the four-loop splitting functions~\cite{Moch:2017uml,Moch:2021qrk}.
	
	\paragraph{Multiloop amplitude construction.}
	Multi-loop scattering amplitudes are typically decomposed in terms of so-called master integrals and their algebraic coefficients.
    This decomposition can be achieved by generating Feynman diagrams, applying projectors to them, and using
    integration-by-parts identities~\cite{Chetyrkin:1981qh} to reduce the resulting 
    integrals to master integrals.
    It is well known that when considering helicity amplitudes their analytic expressions are considerably simpler than what their
    Feynman diagrammatic representation would suggest. This simplicity is made even more evident when employing a specially chosen basis of special functions, which reflects the holomorphic structure of the amplitude.
    
    In order to construct an analytic integrand, projector methods have been proposed recently~\cite{Peraro:2019cjj,Peraro:2020sfm,Chen:2019wyb} that avoid evanescent structures in the calculation of helicity amplitudes in dimensional regularization, which is of particular relevance for higher multiplicities.
    Techniques based on numerical integrand reduction have also been developed at the multi-loop
    level~\cite{Mastrolia:2011pr,Badger:2012dp,Zhang:2012ce,Mastrolia:2016dhn}, which build on the success of the one-loop {\abbrev OPP}
    method~\cite{delAguila:2004nf,Ossola:2006us,Ossola:2007bb}. They allow to efficiently extract integrand coefficients by 
    matching to numerical evaluations. One novel integrand parametrization is that of the \textit{master-surface} decomposition~\cite{Ita:2015tya} 
    which allows to incorporate the reduction to master integrals directly in the integrand matching procedure.
    This parametrization was key for pushing the numerical unitarity method~\cite{Ellis:2007br,Giele:2008ve,Berger:2008sj} to the multi-loop 
    level~\cite{Abreu:2017xsl,Abreu:2017idw,Abreu:2017hqn,Abreu:2018jgq} where full amplitudes can be computed numerically based 
    on tree-level amplitudes, removing the need of building analytic integrands.
	Efforts are also underway to extend successful one-loop numerical techniques based on 4-dimensional integrand reduction and recursive techniques for rational term calculations to the two-loop level~\cite{Pozzorini:2020hkx,Lang:2020nnl,Lang:2021hnw,Pozzorini:2022ohr}.

    A core computational bottleneck in the computation of amplitudes is the handling of linear relations
    between Feynman integrals (integration-by-parts identities).
    In principle, Laporta's algorithm~\cite{Laporta:2000dsw} provides a general solution to the problem.
    Currently, several public reduction codes are available, for example \texttt{Fire}~\cite{Smirnov:2019qkx}, \texttt{Reduze}~\cite{vonManteuffel:2012np}, \texttt{LiteRed}~\cite{Lee:2013mka}, and \texttt{Kira}~\cite{Klappert:2020nbg}.
    In practice, one still faces significant practical challenges for state-of-the-art problems due to the large size
    of the required systems of equations and the complexity of the emerging algebraic coefficients.
    In recent years, a deeper systematic understanding of the linear relations has been gained
    through methods from polynomial ideal theory~\cite{Gluza:2010ws,Larsen:2015ped,Georgoudis:2016wff,Bohm:2017qme,Lee:2014tja,Bitoun:2017nre,Agarwal:2020dye}. 
    In particular, the computation of syzygies allows to systematically reduce integrals to a set
    of master integrals without the introduction of many auxiliary integrals.
    
    Choosing suitable master integrals helps to avoid the introduction of spurious denominator factors
    (singularities)~\cite{Smirnov:2020quc,Usovitsch:2020jrk}, which helps to reduce the algebraic complexity.
    Pioneering work based on intersection theory~\cite{Mastrolia:2018uzb,Frellesvig:2019kgj} raises the question
    whether such completely new techniques to efficiently perform integral reductions for complicated problems may become available in the future.

    \paragraph{Sampling methods and distributed computations.}
	Perhaps one of the most drastic changes in the field regards the wide use of various numerical sampling techniques to construct analytic expressions for multi-loop amplitudes. 
	Since all multi-scale multi-loop amplitudes mentioned
	above are functions of not too many variables (say up to six variables), one might expect that symbolic results allow for faster evaluation compared to numerical approaches.
	However, due to the sheer size of the intermediate algebraic expressions, arriving at these results by symbolic manipulations can become very challenging.
	Instead, numerical or semi-numerical approaches avoid computational complexity at intermediate stages and are better suited for distributed computations.
	
	The use of modular arithmetic (finite fields) in the calculation of integration-by-parts reductions was proposed in ref.~\cite{vonManteuffel:2014ixa}.
	The general idea is to perform computations with different integer samples for the variables in the problem
	and to reconstruct the symbolic information of interest from many such evaluations.
    A major advantage of this approach is that intermediate expression swell can be systematically avoided.
    Soon after it was shown~\cite{Peraro:2016wsq} how helicity amplitude coefficients can be reconstructed from finite field samples, treating multivariate rational functions through a carefully crafted nested approach.
	
	Finite field sampling and functional reconstruction techniques like these have become standard for cutting-edge problems in the past few years, and several public~\cite{Klappert:2019emp,Peraro:2019svx,Smirnov:2019qkx,Klappert:2020nbg,Abreu:2020xvt} and private implementations
	have been developed. These algorithms are rather flexible and can be employed at several stages of the calculations, allowing simplifications to be readily applied before the analytic structures are reconstructed. This includes for example removing lower-loop information by targeting the corresponding \enquote{finite remainder}, accessing the amplitude at the special function level by expanding it as a Laurent series in $\epsilon$,  precomputing the denominator structure of the rational functions involved, among others.
    Another numerical approach for extracting analytic expressions but using high-precision floating-point
	arithmetic together with carefully crafted ansatze was presented in ref.~\cite{Laurentis:2019bjh} (see applications
	in ref.~\cite{Budge:2020oyl,Campbell:2021mlr,Campbell:2022qpq}) and recently extended to an approach closely
	related to finite fields, based on $p$-adic numbers~\cite{DeLaurentis:2022otd}.

    To use an amplitude calculation in a phenomenological application, the corresponding numerical
	evaluation has to be efficient (in time and memory) and numerically stable over phase space even close to singular regions.
    The identification of linear relations between rational functions
    and a suitable representation of the rational functions themselves
    can help to reduce the size and numerical stability of the
    amplitudes.
    In particular, using multivariate partial fraction decomposition~\cite{Pak:2011xt,Raichev:2012pf}
    instead of a common-denominator form can help significantly
    in that regard, see for example ref.~\cite{Abreu:2019odu}.
    Public implementations of recent algorithms in the \texttt{Singular} CAS~\cite{Bendle:2019csk} and in the \texttt{MultivariateApart}~\cite{Heller:2021qkz} package can be used for distributed computation
    of the necessary manipulations in a cluster-friendly way.
	
	These novel techniques have opened the door to perform computer algebraic computations for quantum field theory on \HPC{} systems. These novel methods typically use integer arithmetic rather than floating point arithmetic and, 
	depending on the problem, they may require a significant amount of memory per core.
    The development of (public) codes for challenging amplitude calculations in a \HPC{} friendly way is an ongoing effort.
    Currently, in some situations, one may need to resort to implementations that have large memory demands and require
    run-times that are hard to predict and possibly well beyond available batch limits on a given cluster.
	
\section{Applications in collider phenomenology}\label{sec:applications}

    Ultimately multi-loop amplitudes and other associated ingredients enter the calculation of cross sections and other observables.
        To obtain cross sections, degenerate final states at each order in perturbation theory must be summed over. This in turn requires the introduction of a mechanism to regularize and cancel related infrared (\IR{}) divergences, the so-called \IR{} subtraction methods. 
        
\paragraph{\IR{} subtraction methods.}
    
    Subtraction methods are available that cancel \IR{} singularities 
    locally in phase space, or that are based on a single \IR{} cutoff (slicing) that must be extrapolated to a vanishing cutoff size. In general, the complexity of subtraction methods increases with the number of colored particles until all different singular limits of a given perturbative order are probed and extensions to more particles turn into a combinatoric problem. For example in \NNLO{} three-jet production all double-singular limits are probed among colored particles.
    
    While local subtraction schemes require significant development time, they generally lead to Monte Carlo phase 
    space integrations that are numerically easier to perform. Slicing methods on the other hand are 
    simpler to construct and have the benefit that they can easily re-use lower order components. 
    An example slicing method is the one based on factorization in the $q_T$ variable, which is currently 
    the only technique used to obtain fully differential \NNNLO{} results for a color singlet 
    process~\cite{Chen:2022cgv}. If a fully inclusive \NNNLO{} calculation is available, also projection to 
    Born ({\abbrev P2B}) is a method to perform an efficient and fully differential
    calculation~\cite{Cacciari:2015jma}.

	Slicing methods are typically based on factorization in a kinematic observable. At \NNLO{} methods are available 
	based on the $q_T$ observable~\cite{Catani:2007vq,Becher:2010tm,Billis:2019vxg} for color singlet systems or based on $N$-jettiness~\cite{Stewart:2010tn,Boughezal:2015dva,Gaunt:2015pea} for generic processes. With 
	$N$-jettiness the ingredients are available for \NNLO{} calculations with up to one colored final-state 
	particle. \NNNLO{} (three loop) soft~\cite{Li:2016ctv,Vladimirov:2016dll} and beam functions~\cite{Luo:2020epw,Ebert:2020yqt} in $q_T$ factorization allow for fully differential \NNNLO{} calculations of colorless final states, see also the Snowmass white paper on \enquote{The Path forward to \NNNLO{}}~\cite{Caola:2022ayt}. The calculation of power corrections allows for accelerated calculations using larger slicing cutoffs, see e.g.~\cite{Moult:2016fqy,Boughezal:2018mvf,Ebert:2018gsn}. A different slicing approach at \NNLO{} has been undertaken in ref.~\cite{Herzog:2018ily}.
	
	As for local subtraction schemes, Antenna subtractions~\cite{Gehrmann-DeRidder:2005btv,Currie:2013vh} and sector-improved residue subtraction 
	(\STRIPPER{})~\cite{Czakon:2010td,Boughezal:2011jf,Czakon:2014oma} have been applied up to two \cite{Currie:2017eqf,Gehrmann-DeRidder:2019ibf,Czakon:2019tmo} and three \cite{Czakon:2021mjy,Chen:2022ktf} colored final states in $pp$ collisions.
	Nested soft collinear subtractions \cite{Caola:2017dug} have been applied to deep inelastic scattering ({\abbrev DIS}), 
	Higgs production and mixed \QCDEW{} corrections in Drell-Yan 
	production~\cite{Asteriadis:2019dte,Asteriadis:2021gpd,Buccioni:2022kgy}, {\abbrev ColorFulNNLO}~\cite{Somogyi:2005xz} 
	to $e^+e^-$ three-jet production~\cite{DelDuca:2016ily} and $H\to b\bar{b}$~\cite{DelDuca:2015zqa}, and local analytic subtraction~\cite{Magnea:2018hab} to Drell-Yan production. A recent comprehensive overview of all past processes implemented using the various subtraction schemes is available in ref.~\cite{Heinrich:2020ybq}, table 5. The local subtraction methods differ in the partitioning of phase-space and analyticity of counterterms. Extensions beyond \NNLO{} will be a major undertaking both in phase-space partitioning and integration of triple singular counterterms and 
	one can expect progress on them during the next decade. Overall the current development focus is still at the level of refining, automating and generalizing current methods at \NNLO{}.
	
	The current frontier in multi-loop phenomenology revolves around \NNLO{} \QCD{} corrections for  $2\to3$ processes with massless particles, two-loop mixed \QCDEW{} corrections for $2\to 1$ and $2\to2$ processes, corrections to $2\to2$ processes involving two-loop amplitudes with additional 
	internal masses (massive \QCD{}) and fiducial \NNNLO{} \QCD{} $2\to1$ processes. 
	For one-loop phenomenology, high-multiplicity processes play a special role as they also consume a considerable
	amount of resources. In the following we list a few examples of state-of-the-art applications that rely on the 
	advancements presented in this white paper.
	
	\paragraph{$2\to 1$ fiducial \NNNLO{}.}
	    Currently, the upper end of computational resource requirements is consumed by fiducial \NNNLO{} calculations.
	    In the $q_T$ subtraction approach, \NNLO{} real emissions need to be determined with high precision in regions close to singularities, resulting in significant computational costs of up to about 10M \CPU{} core hours.
	    
	    Fiducial Drell-Yan production at \NNNLO{} has been computed in ref.~\cite{Chen:2022cgv} using $q_T$ subtractions and using antenna subtractions for the \NNLO{} real emission process. Fiducial Higgs results have been presented at \NNNLO{} \cite{Chen:2021isd} using {\abbrev P2B} and using antenna subtractions for the real emission, reducing resource requirements by an order of magnitude or two compared to slicing. Improved Higgs results that include $q_T$ resummation of fiducial power corrections at the level of \NNNLO{} have been presented in ref.~\cite{Billis:2021ecs}.
	
	\paragraph{$2\to 3$ at \NNLO{}.} 
	
	    Phenomenology for $2\to 3$ processes at \NNLO{} broadly uses a few hundred thousand up to a million 
	    \CPU{} core hours per project, depending on the final state.
	    In this, the real emission is a major factor. While the analytic calculation of the two-loop amplitudes may need 
	    substantial computational resources by itself, the numerical evaluation of carefully optimized representations 
	    typically does not represent a major challenge. 
	    Alternatively, the virtual corrections have been integrated using a pre-computed interpolation grid.
	    
       Triphoton production ($\gamma\gamma\gamma$) has been computed within the \STRIPPER{} 
       framework~\cite{Chawdhry:2019bji}. It has also been studied with $q_T$ subtractions \cite{Kallweit:2020gcp}, 
       evaluating the two-loop integrals with the \texttt{PentagonFunctions++} library~\cite{Chicherin:2020oor} using
       the two-loop matrix elements of ref.~\cite{Abreu:2020cwb}.
       Using \STRIPPER{}, a calculation of $3$-jet production at \NNLO{} has been presented~\cite{Czakon:2021mjy} using 
       the double-virtual contributions of ref.~\cite{Abreu:2021oya}
       (see also recent related work in ref.~\cite{Chen:2022ktf}). Using the same framework also 
       $\gamma\gamma+$jet production at \NNLO{} has been computed \cite{Chawdhry:2021hkp} employing the double-virtual
       contributions of refs.~\cite{Chawdhry:2021mkw,Agarwal:2021vdh}. For the latter process also the gluon-initiated 
       (loop-induced) contributions have been computed including \NLO{} corrections~\cite{Badger:2021ohm} employing the
       two-loop matrix elements of ref.~\cite{Badger:2021imn}.

	\paragraph{$2\to 1$ and $2\to 2$ with many scales.}
	
       The computational requirements for $2\to 2$ color singlet processes at \NNLO{} without internal masses are much lower, at the order of just a few hundred \CPU{} core hours.
       However, the demands increase drastically in the presence of jets due to the more complicated singularity structure, in some cases reaching a few hundred thousand \CPU{} core hours, see e.g.\  refs.~\cite{Asteriadis:2021gpd,Pellen:2021vpi,Czakon:2020qbd,Czakon:2021ohs,Currie:2017eqf,Gehrmann-DeRidder:2019ibf}. Computational demands can also increase due to cuts for otherwise simple processes \cite{Alekhin:2021xcu}.

	    The calculational and computational complexity also increases with a larger number of scales, as for example in \EW{} corrections or in \QCD{} with massive quarks.
	    Often such problems are at the frontier of loop-integral complexity.
	    In some cases, such as for the mixed \QCDEW{} corrections to Drell-Yan processes,
	    analytical representations of two-loop amplitudes with internal masses were still an option~\cite{Behring:2020cqi,Heller:2020owb,Buccioni:2022kgy},
	    rendering the computational demands for phenomenological applications unproblematic.
	    In other cases, numerical approaches are the only available option.
	    The computational demands of the latter are up to a few hundred thousand \CPU{} core hours, while memory requirements are again dominated by the reduction to master integrals and can be especially demanding here, up to 2TB per node for some surveyed calculations.
	    
	   For example, for $H$+jet and di-Higgs production the full top-quark mass dependence has been obtained 
	   only numerically with sector decomposition~\cite{Jones:2018hbb,Borowka:2016ehy,Borowka:2016ypz}. These calculations 
	   used around 10,000 hours on \GPU{}s (generation 2012) each for the numerical integration, with a median of two hours 
	   (and up to two days) necessary for evaluating one phase space point.
	   These calculations are the only ones in our survey that systematically use \GPU{} resources. Given the development
	   of this type of technology in most planned \HPC{} systems, such applications may increase in the future.

	   The numerical evaluation of loop integrals through series solutions of differential equations is becoming increasingly popular and has been used in several phenomenological applications already. In ref.~\cite{Czakon:2021yub} the exact 
	   top-mass dependence in fully inclusive Higgs production has been calculated this way at \NNLO{} in \QCD{}. The mixed \QCDEW{} corrections to Drell-Yan production presented in ref.~\cite{Bonciani:2021zzf} relied 
	   on numerical evaluations of the most complicated integrals through series expansions.
	   A further application of the method was the calculation of two-loop non-factorizable \NNLO{} \QCD{} contributions in single-top-quark production \cite{Bronnum-Hansen:2021pqc}. 
    
    \paragraph{High-multiplicity \NLO{} and \NNLO{} matched to parton showers.}
    
        State-of-the-art \NLO{} predictions for processes with high multiplicities (see e.g.\ $t\bar{t} b\bar{b}$ production 
        studies~\cite{Bevilacqua:2021cit,Bevilacqua:2022twl,Denner:2020orv}) can have computational 
        costs similar to current state-of-the-art \NNLO{} calculations, at the order of 100k--1M \CPU{} core hours.

        Resource demands for event generators that include parton shower, hadronization and possibly 
        detector simulation are overall orders of magnitude larger than parton-level predictions. They 
        fall outside the scope of this white paper, see e.g.\ the Snowmass white paper on event generators \cite{Campbell:2022qmc}. Nevertheless, we would like to mention that meanwhile
        \NNLO{} calculations are matched to parton showers, and the resource requirements of matched 
        processes reach those of more complicated \NNLO{} calculations, at the level of a few hundred 
        thousand \CPU{} core hours, see e.g.\ refs.~\cite{Mazzitelli:2021mmm,Lombardi:2020wju,Cridge:2021hfr,Alioli:2021egp}.
       
	\begin{figure}
	    \centering
	    \includegraphics[width=0.5\textwidth]{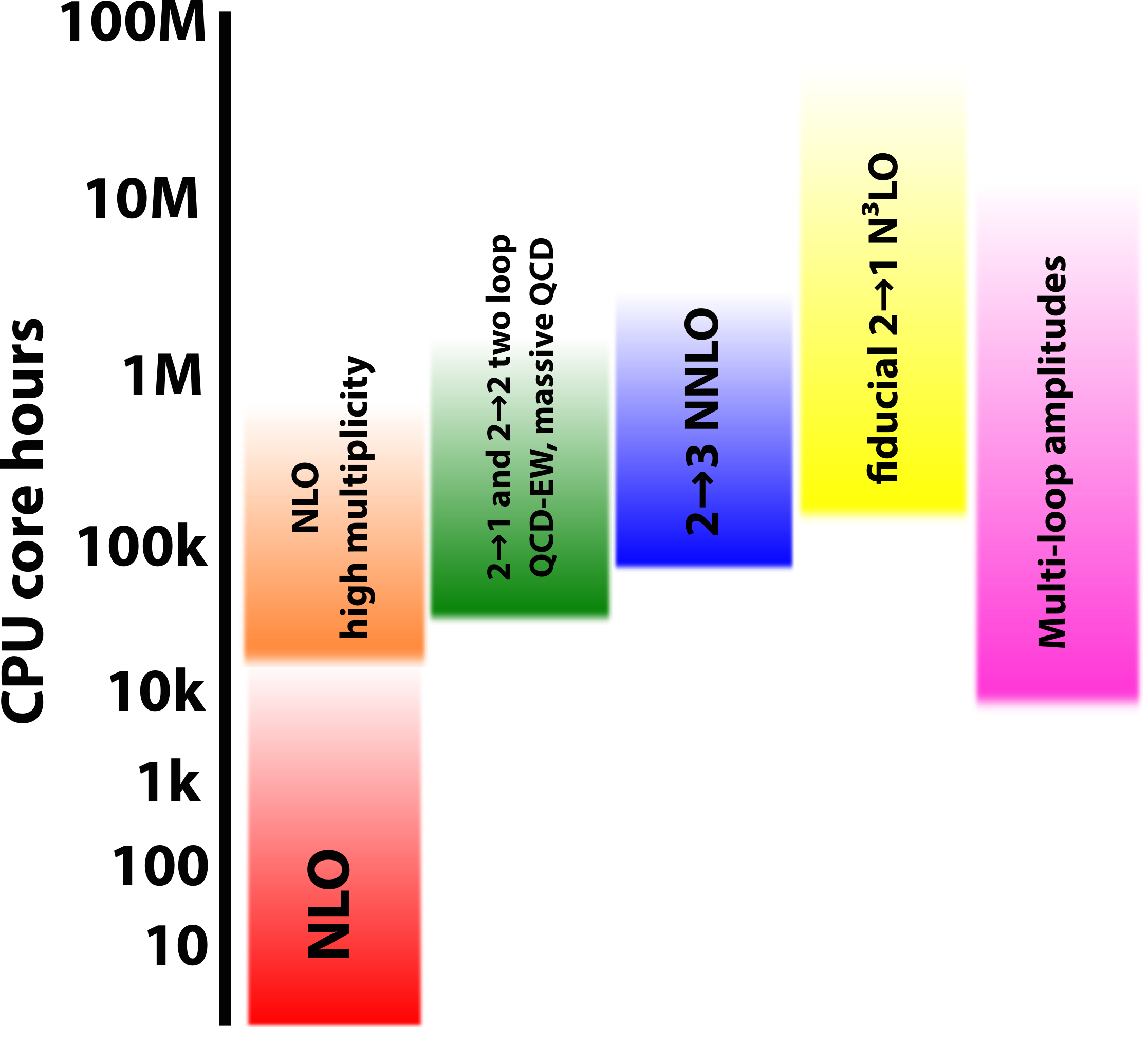}
	    \caption{Run-time requirements of recent perturbative calculations for collider phenomenology.
	    Memory requirements ranged up to about 2 TB of RAM per node.}
	    \label{fig:corehours}
	\end{figure}
	
	\section{Conclusions and outlook}\label{sec:outlook}

	During the last ten years tremendous progress on higher-order calculations for collider physics has been achieved. The 
	use of \HPC{} systems for these applications is now both a standard as well as a requirement. With Snowmass 2013, the computational requirements of perturbative calculations were assessed and several 
	questions were posed for discussion in order to efficiently push the boundaries of precision studies~\cite{Campbell:2013qaa,Hoche:2013zja}. For this Snowmass 2021 
	white paper we provide an update regarding the current status of the field and give an outlook for the decade ahead.
	
	\paragraph{Use of \CPU{} parallel computing.}
	When comparing previous with current state-of-the-art computations, we notice that node hour requirements are not \emph{vastly} different despite a very significant increase in the complexity of the treated problems.
	In many cases, theoretical developments lead to much more efficient approaches.
	In addition, single core performance has increased by a factor of four to five since 2013 and single \CPU{}/node performance has increased by a factor of 25-30 for state-of-the-art hardware. This discrepancy between single core performance and \CPU{}/node performance is a trend that is expected to continue in the future, given that transistors already reach a few nm in size.  
	A consequence of this is that using parallel high performance computing becomes ever more important to benefit from improvements in computing technology.
	
	Parallel computing is used by the community, although mostly in terms of independent jobs and not with inter-process communication using for example MPI. For some cases this is sufficient or advantageous, but in other cases, e.g.\ large phase-space integrations, calculations can benefit from efforts in using distributed computations based on MPI or a similar framework.
	
	\paragraph{Use of \GPU{}s.}
	Modern computing paradigms focus more and more on the use of specialized \GPU{}s instead of general purpose \CPU{}s.
    All new leading {\abbrev DOE} clusters like NERSC Perlmutter, Argonne Aurora and Oak Ridge Frontier follow this paradigm and focus on nodes with \GPU{}s. 
    \GPU{}s offer distinct advantages for single instruction multiple thread and multiple data problems that do not require irregular access to large amounts of memory.
    The
    use of \GPU{}s is still unclear in our field, since many problems in our field rely on the numerical evaluation or handling of algebraic expressions which are large and/or require irregular memory access patterns. So far, \GPU{}s have found application to cutting-edge problems with the numerical integration of sector decomposed loop integrals. A first step for future applications could include the efficient evaluation of one-loop amplitudes. This would help the huge computational requirements for \NLO{} high-multiplicity evaluations, but also for the real emission integrations for \NNLO{} calculations. Since the efficient use of \GPU{}s is still unclear, future computing for our community will still need to focus on providing \CPU{} resources without attached \GPU{} resources.

	\paragraph{Memory and run-times requirements.}
	A number of multi-loop amplitude calculations involving computer algebra are limited by the available memory per core or even per node. In addition, long run-times may be required and break-pointing might not be 
	a good option in practice due the required additional development efforts or the use of proprietary software and external dependencies.
	We note that both memory and run-time requirements may be intrinsically difficult to predict for such type of problems.
	Moreover, often a variety of different approaches and codes will be tested or combined, and only limited human resources are available for software development to adapt to the given constraints on the available clusters.
	Currently, many of these calculations are being performed on local computing resources, partly outside of a larger shared cluster infrastructure to circumvent constraints imposed by general cluster policies.
	Regarding availability of resources, more high-memory nodes and more flexibility on long job run-times could provide effective help.
	Through snapshots, virtualization solutions in cluster environments may resolve the tension between such calculational demands and requirements of cluster maintenance.

	\paragraph{Machine learning.}
	Machine learning, while being explored, has not made significant impact yet within the research scope presented here at the multi-loop level. Machine learning is being explored for improving traditional optimization and interpolation problems. For example the optimization of phase-space integration has been studied, as well as the fast interpolation of multi-loop hard scattering functions. So far the most promising applications focus on the event generation beyond the parton level, see e.g.\ the Snowmass white paper on \enquote{Machine Learning and \LHC{} Event Generation} \cite{Butter:2022rso}.

	\paragraph{Availability and usability of codes.}
    The public availability of codes and results is a crucial aspect to improve efficiency of the community and to reduce friction in exchanges and comparisons. State-of-the-art multi-loop amplitudes are nowadays provided by many authors in machine-readable format, and increasingly also building blocks like master integrals or even reduction tables.
    The level of sophistication of modern tools for tasks like integral reduction or integral evaluation clearly motivates their public availability, such that they can be used by different research groups.
    Indeed, an increasing number of authors invest the required time to make their codes accessible to others and share them early on.
    We believe this is an important development which increases productivity in our field,
    and that attention should be payed to longer-term career perspectives of young researchers engaging in these efforts.
    Of course, public codes with proper documentation also avoids loss of intellectual achievements if Ph.D.\ students or postdocs leave the field.
    Public availability is also relevant for entire cross section calculations 
    to allow for derived work and comparative studies.
    Currently, a subset of \NNLO{} codes are publicly available.

	Codes using proprietary software like Mathematica or Maple are limited by license availability for large deployments on clusters.
	In practice, such systems may offer distinct advantages as a development platform and are therefore frequently used, both for less resource-critical tasks and for prototyping.
	While limitations regarding the usage on clusters can eventually be avoided with dedicated implementations not relying on proprietary software, it remains a problem-specific decision whether the expected gains justify this additional effort.

	\paragraph{Grade of automation.}
	Nowadays, highly automated codes are available for the computations of \NLO{} corrections, including even electroweak interactions.
	While 2013 marked the early advent of \NNLO{} calculations, meanwhile a much higher level of sophistication has been achieved by treating also $2\to 3$ processes like $pp\to 3$ jets and lower multiplicity processes with many scales at \NNLO{}. Amplitudes, loop integrals and \IR{} subtractions all pose highly difficult problems that require dedicated efforts for their automation.
	With current technology we are indeed getting closer to the automation of \NNLO{} calculations -- raw ingredients are available for amplitudes, the numerical evaluation of loop integrals and \NNLO{} \IR{} subtractions to high multiplicities.
	Typically generic numerical methods allow for easier automation, while targeted analytical work can provide particularly efficient implementations.
	Overall, a lot of consolidation work is still necessary for automated \NNLO{} frameworks, which we expect to take place in the upcoming decade.

	\paragraph{Closing remarks.}
	
	Overall the sample state-of-the-art projects surveyed for this white paper took each 2-5 years of total PhD/postdoc research time. This is a significant time in terms of the average hiring span of PhD students and postdocs and points to the trend for larger collaborations to exploit synergies. Furthermore, each 
	precision calculation (at \NNLO{} or beyond) relies on more than a decade of developments in the \IR{} subtraction frameworks and in the mathematical and algorithmic developments for amplitudes and loop integrals.
	This motivates a continued effort to provide public and well documented codes for tools and predictions, so that the whole theory and experimental communities can benefit from state-of-the-art advancements.
	
	Today's typical computational resource requirements for phenomenological predictions are at the order of a few hundred thousand core hours and reach up to about 10M core hours; 
	a distribution of them is sketched in \cref{fig:corehours}. This is in stark contrast to the last Snowmass planning in 2013, where the use of cluster resources was more of an exception rather than standard.
	The majority of people answering our survey deemed the currently-available resources sufficient for their current projects, although in some cases easier access to suitable cluster resources would have been beneficial.
	For the next 5-15 years, we expect to see \NNLO{} predictions for more complex final states, the automation of these calculations, and \NNNLO{} predictions for diboson processes.
	Numerical and semi-numerical methods as well as distributed computing are expected to play an essential role in these efforts.
	It remains unclear to which degree \GPU{}s can be employed, such that the availability of 
	\CPU{}-oriented cluster resources stays important.
    Nodes with multiple TB of memory continue to be of high relevance for computer-algebraic components of the calculations.
    Flexible job run times significantly benefit some developments, and virtualization solutions with snapshot capabilities could enable this in cluster environments.
    
	\paragraph{Acknowledgments.}
	
	We would like to thank the following people for filling out our survey and providing valuable input on the computational resources of their projects: Samuel Abreu, Bakul Agarwal, Konstantin Asteriadis, Simon Badger, Matteo Becchetti, Marco Bonetti, Federico Buccioni, Luca Buonocore, Fabrizio Caola, Gudrun Heinrich, Alexander Huss, Stephen P. Jones, Stefan Kallweit,
	Matthias Kerner, Matteo Marcoli, Javier Mazzitelli, Johannes Michel, Sven Moch, Marco Niggetiedt, Costas Papadopoulos, Mathieu Pellen, Rene Poncelet, Jérémie Quarroz, Luca Rottoli, Gabor Somogyi, Qian Song, Vasily Sotnikov, Matthias Steinhauser, Gherardo Vita, Chen-Yu Wang, Stefan Weinzierl, Marius Wiesemann, Malgorzata Worek, Tongzhi Yang and YuJiao Zhu.

	The work of Fernando Febres Cordero is supported in part by the United States Department of Energy under grant DE-SC0010102.
	Andreas von Manteuffel is supported in part by the National Science Foundation under Grant 2013859.
	Tobias Neumann is supported by the United States Department of Energy under Grant Contract DE-SC0012704.

	\bibliographystyle{JHEP}
	\bibliography{references}

\providecommand{\href}[2]{#2}\begingroup\raggedright\begin{thebibliography}{100}

\bibitem{Hoche:2013zja}
S.~Hoche et~al., \emph{{Working Group Report: Computing for Perturbative QCD}},
   in \emph{{Community Summer Study 2013}: {Snowmass on the Mississippi}}, 9,
  2013, \href{https://arxiv.org/abs/1309.3598}{{\ttfamily 1309.3598}}.

\bibitem{Campbell:2013qaa}
J.~M. Campbell et~al., \emph{{Working Group Report: Quantum Chromodynamics}},
  in \emph{{Community Summer Study 2013}: {Snowmass on the Mississippi}}, 10,
  2013, \href{https://arxiv.org/abs/1310.5189}{{\ttfamily 1310.5189}}.

\bibitem{Bern:2013gka}
Z.~Bern, L.~J. Dixon, F.~Febres~Cordero, S.~H\"oche, H.~Ita, D.~A. Kosower
  et~al., \emph{{Next-to-Leading Order $W + 5$-Jet Production at the LHC}},
  \href{https://doi.org/10.1103/PhysRevD.88.014025}{\emph{Phys. Rev. D}
  {\bfseries 88} (2013) 014025}
  [\href{https://arxiv.org/abs/1304.1253}{{\ttfamily 1304.1253}}].

\bibitem{Melnikov:2006di}
K.~Melnikov and F.~Petriello, \emph{{The $W$ boson production cross section at
  the LHC through $O(\alpha^2_s)$}},
  \href{https://doi.org/10.1103/PhysRevLett.96.231803}{\emph{Phys. Rev. Lett.}
  {\bfseries 96} (2006) 231803}
  [\href{https://arxiv.org/abs/hep-ph/0603182}{{\ttfamily hep-ph/0603182}}].

\bibitem{Li:2012wna}
Y.~Li and F.~Petriello, \emph{{Combining QCD and electroweak corrections to
  dilepton production in FEWZ}},
  \href{https://doi.org/10.1103/PhysRevD.86.094034}{\emph{Phys. Rev. D}
  {\bfseries 86} (2012) 094034}
  [\href{https://arxiv.org/abs/1208.5967}{{\ttfamily 1208.5967}}].

\bibitem{Anastasiou:2005qj}
C.~Anastasiou, K.~Melnikov and F.~Petriello, \emph{{Fully differential Higgs
  boson production and the di-photon signal through next-to-next-to-leading
  order}}, \href{https://doi.org/10.1016/j.nuclphysb.2005.06.036}{\emph{Nucl.
  Phys. B} {\bfseries 724} (2005) 197}
  [\href{https://arxiv.org/abs/hep-ph/0501130}{{\ttfamily hep-ph/0501130}}].

\bibitem{Barnreuther:2012wtj}
P.~B\"arnreuther, M.~Czakon and A.~Mitov, \emph{{Percent Level Precision
  Physics at the Tevatron: First Genuine NNLO QCD Corrections to $q \bar{q} \to
  t \bar{t} + X$}},
  \href{https://doi.org/10.1103/PhysRevLett.109.132001}{\emph{Phys. Rev. Lett.}
  {\bfseries 109} (2012) 132001}
  [\href{https://arxiv.org/abs/1204.5201}{{\ttfamily 1204.5201}}].

\bibitem{Czakon:2013goa}
M.~Czakon, P.~Fiedler and A.~Mitov, \emph{{Total Top-Quark Pair-Production
  Cross Section at Hadron Colliders Through $O(\alpha^4_S)$}},
  \href{https://doi.org/10.1103/PhysRevLett.110.252004}{\emph{Phys. Rev. Lett.}
  {\bfseries 110} (2013) 252004}
  [\href{https://arxiv.org/abs/1303.6254}{{\ttfamily 1303.6254}}].

\bibitem{Gehrmann-DeRidder:2013uxn}
A.~Gehrmann-De~Ridder, T.~Gehrmann, E.~W.~N. Glover and J.~Pires, \emph{{Second
  order QCD corrections to jet production at hadron colliders: the all-gluon
  contribution}},
  \href{https://doi.org/10.1103/PhysRevLett.110.162003}{\emph{Phys. Rev. Lett.}
  {\bfseries 110} (2013) 162003}
  [\href{https://arxiv.org/abs/1301.7310}{{\ttfamily 1301.7310}}].

\bibitem{Boughezal:2013uia}
R.~Boughezal, F.~Caola, K.~Melnikov, F.~Petriello and M.~Schulze, \emph{{Higgs
  boson production in association with a jet at next-to-next-to-leading order
  in perturbative QCD}},
  \href{https://doi.org/10.1007/JHEP06(2013)072}{\emph{JHEP} {\bfseries 06}
  (2013) 072} [\href{https://arxiv.org/abs/1302.6216}{{\ttfamily 1302.6216}}].

\bibitem{Campbell:2019dru}
J.~Campbell and T.~Neumann, \emph{{Precision Phenomenology with MCFM}},
  \href{https://doi.org/10.1007/JHEP12(2019)034}{\emph{JHEP} {\bfseries 12}
  (2019) 034} [\href{https://arxiv.org/abs/1909.09117}{{\ttfamily
  1909.09117}}].

\bibitem{Weinzierl:2022eaz}
S.~Weinzierl, \emph{{Feynman Integrals}},
  \href{https://arxiv.org/abs/2201.03593}{{\ttfamily 2201.03593}}.

\bibitem{Heller:2019gkq}
M.~Heller, A.~von Manteuffel and R.~M. Schabinger, \emph{{Multiple
  polylogarithms with algebraic arguments and the two-loop EW-QCD Drell-Yan
  master integrals}},
  \href{https://doi.org/10.1103/PhysRevD.102.016025}{\emph{Phys. Rev. D}
  {\bfseries 102} (2020) 016025}
  [\href{https://arxiv.org/abs/1907.00491}{{\ttfamily 1907.00491}}].

\bibitem{Moriello:2019yhu}
F.~Moriello, \emph{{Generalised power series expansions for the elliptic planar
  families of Higgs + jet production at two loops}},
  \href{https://doi.org/10.1007/JHEP01(2020)150}{\emph{JHEP} {\bfseries 01}
  (2020) 150} [\href{https://arxiv.org/abs/1907.13234}{{\ttfamily
  1907.13234}}].

\bibitem{Agarwal:2020dye}
B.~Agarwal, S.~P. Jones and A.~von Manteuffel, \emph{{Two-loop helicity
  amplitudes for $gg \to ZZ$ with full top-quark mass effects}},
  \href{https://doi.org/10.1007/JHEP05(2021)256}{\emph{JHEP} {\bfseries 05}
  (2021) 256} [\href{https://arxiv.org/abs/2011.15113}{{\ttfamily
  2011.15113}}].

\bibitem{Bronnum-Hansen:2020mzk}
C.~Br\o{}nnum-Hansen and C.-Y. Wang, \emph{{Contribution of third generation
  quarks to two-loop helicity amplitudes for W boson pair production in gluon
  fusion}}, \href{https://doi.org/10.1007/JHEP01(2021)170}{\emph{JHEP}
  {\bfseries 01} (2021) 170}
  [\href{https://arxiv.org/abs/2009.03742}{{\ttfamily 2009.03742}}].

\bibitem{Bronnum-Hansen:2021olh}
C.~Br\o{}nnum-Hansen and C.-Y. Wang, \emph{{Top quark contribution to two-loop
  helicity amplitudes for $Z$ boson pair production in gluon fusion}},
  \href{https://doi.org/10.1007/JHEP05(2021)244}{\emph{JHEP} {\bfseries 05}
  (2021) 244} [\href{https://arxiv.org/abs/2101.12095}{{\ttfamily
  2101.12095}}].

\bibitem{Bronnum-Hansen:2021pqc}
C.~Br\o{}nnum-Hansen, K.~Melnikov, J.~Quarroz and C.-Y. Wang, \emph{{On
  non-factorisable contributions to t-channel single-top production}},
  \href{https://doi.org/10.1007/JHEP11(2021)130}{\emph{JHEP} {\bfseries 11}
  (2021) 130} [\href{https://arxiv.org/abs/2108.09222}{{\ttfamily
  2108.09222}}].

\bibitem{Papadopoulos:2015jft}
C.~G. Papadopoulos, D.~Tommasini and C.~Wever, \emph{{The Pentabox Master
  Integrals with the Simplified Differential Equations approach}},
  \href{https://doi.org/10.1007/JHEP04(2016)078}{\emph{JHEP} {\bfseries 04}
  (2016) 078} [\href{https://arxiv.org/abs/1511.09404}{{\ttfamily
  1511.09404}}].

\bibitem{Gehrmann:2018yef}
T.~Gehrmann, J.~M. Henn and N.~A. Lo~Presti, \emph{{Pentagon functions for
  massless planar scattering amplitudes}},
  \href{https://doi.org/10.1007/JHEP10(2018)103}{\emph{JHEP} {\bfseries 10}
  (2018) 103} [\href{https://arxiv.org/abs/1807.09812}{{\ttfamily
  1807.09812}}].

\bibitem{Abreu:2020jxa}
S.~Abreu, H.~Ita, F.~Moriello, B.~Page, W.~Tschernow and M.~Zeng,
  \emph{{Two-Loop Integrals for Planar Five-Point One-Mass Processes}},
  \href{https://doi.org/10.1007/JHEP11(2020)117}{\emph{JHEP} {\bfseries 11}
  (2020) 117} [\href{https://arxiv.org/abs/2005.04195}{{\ttfamily
  2005.04195}}].

\bibitem{Canko:2020ylt}
D.~D. Canko, C.~G. Papadopoulos and N.~Syrrakos, \emph{{Analytic representation
  of all planar two-loop five-point Master Integrals with one off-shell leg}},
  \href{https://doi.org/10.1007/JHEP01(2021)199}{\emph{JHEP} {\bfseries 01}
  (2021) 199} [\href{https://arxiv.org/abs/2009.13917}{{\ttfamily
  2009.13917}}].

\bibitem{Chicherin:2020oor}
D.~Chicherin and V.~Sotnikov, \emph{{Pentagon Functions for Scattering of Five
  Massless Particles}},
  \href{https://doi.org/10.1007/JHEP12(2020)167}{\emph{JHEP} {\bfseries 20}
  (2020) 167} [\href{https://arxiv.org/abs/2009.07803}{{\ttfamily
  2009.07803}}].

\bibitem{Badger:2021nhg}
S.~Badger, H.~B. Hartanto and S.~Zoia, \emph{{Two-Loop QCD Corrections to
  Wbb\textasciimacron{} Production at Hadron Colliders}},
  \href{https://doi.org/10.1103/PhysRevLett.127.012001}{\emph{Phys. Rev. Lett.}
  {\bfseries 127} (2021) 012001}
  [\href{https://arxiv.org/abs/2102.02516}{{\ttfamily 2102.02516}}].

\bibitem{Chicherin:2021dyp}
D.~Chicherin, V.~Sotnikov and S.~Zoia, \emph{{Pentagon functions for one-mass
  planar scattering amplitudes}},
  \href{https://doi.org/10.1007/JHEP01(2022)096}{\emph{JHEP} {\bfseries 01}
  (2022) 096} [\href{https://arxiv.org/abs/2110.10111}{{\ttfamily
  2110.10111}}].

\bibitem{Abreu:2021smk}
S.~Abreu, H.~Ita, B.~Page and W.~Tschernow, \emph{{Two-loop hexa-box integrals
  for non-planar five-point one-mass processes}},
  \href{https://doi.org/10.1007/JHEP03(2022)182}{\emph{JHEP} {\bfseries 03}
  (2022) 182} [\href{https://arxiv.org/abs/2107.14180}{{\ttfamily
  2107.14180}}].

\bibitem{Papadopoulos:2019iam}
C.~G. Papadopoulos and C.~Wever, \emph{{Internal Reduction method for computing
  Feynman Integrals}},
  \href{https://doi.org/10.1007/JHEP02(2020)112}{\emph{JHEP} {\bfseries 02}
  (2020) 112} [\href{https://arxiv.org/abs/1910.06275}{{\ttfamily
  1910.06275}}].

\bibitem{Kardos:2022tpo}
A.~Kardos, C.~G. Papadopoulos, A.~V. Smirnov, N.~Syrrakos and C.~Wever,
  \emph{{Two-loop non-planar hexa-box integrals with one massive leg}},
  \href{https://arxiv.org/abs/2201.07509}{{\ttfamily 2201.07509}}.

\bibitem{Henn:2020lye}
J.~Henn, B.~Mistlberger, V.~A. Smirnov and P.~Wasser, \emph{{Constructing d-log
  integrands and computing master integrals for three-loop four-particle
  scattering}}, \href{https://doi.org/10.1007/JHEP04(2020)167}{\emph{JHEP}
  {\bfseries 04} (2020) 167}
  [\href{https://arxiv.org/abs/2002.09492}{{\ttfamily 2002.09492}}].

\bibitem{Canko:2021xmn}
D.~D. Canko and N.~Syrrakos, \emph{{Planar three-loop master integrals for $2
  \to 2$ processes with one external massive particle}},
  \href{https://arxiv.org/abs/2112.14275}{{\ttfamily 2112.14275}}.

\bibitem{Henn:2016men}
J.~M. Henn, A.~V. Smirnov, V.~A. Smirnov and M.~Steinhauser, \emph{{A planar
  four-loop form factor and cusp anomalous dimension in QCD}},
  \href{https://doi.org/10.1007/JHEP05(2016)066}{\emph{JHEP} {\bfseries 05}
  (2016) 066} [\href{https://arxiv.org/abs/1604.03126}{{\ttfamily
  1604.03126}}].

\bibitem{Henn:2019rmi}
J.~M. Henn, T.~Peraro, M.~Stahlhofen and P.~Wasser, \emph{{Matter dependence of
  the four-loop cusp anomalous dimension}},
  \href{https://doi.org/10.1103/PhysRevLett.122.201602}{\emph{Phys. Rev. Lett.}
  {\bfseries 122} (2019) 201602}
  [\href{https://arxiv.org/abs/1901.03693}{{\ttfamily 1901.03693}}].

\bibitem{Henn:2019swt}
J.~M. Henn, G.~P. Korchemsky and B.~Mistlberger, \emph{{The full four-loop cusp
  anomalous dimension in $\mathcal{N}=4$ super Yang-Mills and QCD}},
  \href{https://doi.org/10.1007/JHEP04(2020)018}{\emph{JHEP} {\bfseries 04}
  (2020) 018} [\href{https://arxiv.org/abs/1911.10174}{{\ttfamily
  1911.10174}}].

\bibitem{vonManteuffel:2019gpr}
A.~von Manteuffel and R.~M. Schabinger, \emph{{Planar master integrals for
  four-loop form factors}},
  \href{https://doi.org/10.1007/JHEP05(2019)073}{\emph{JHEP} {\bfseries 05}
  (2019) 073} [\href{https://arxiv.org/abs/1903.06171}{{\ttfamily
  1903.06171}}].

\bibitem{vonManteuffel:2020vjv}
A.~von Manteuffel, E.~Panzer and R.~M. Schabinger, \emph{{Cusp and collinear
  anomalous dimensions in four-loop QCD from form factors}},
  \href{https://doi.org/10.1103/PhysRevLett.124.162001}{\emph{Phys. Rev. Lett.}
  {\bfseries 124} (2020) 162001}
  [\href{https://arxiv.org/abs/2002.04617}{{\ttfamily 2002.04617}}].

\bibitem{Agarwal:2021zft}
B.~Agarwal, A.~von Manteuffel, E.~Panzer and R.~M. Schabinger, \emph{{Four-loop
  collinear anomalous dimensions in QCD and $\mathcal{N} = 4$ super
  Yang-Mills}},  \href{https://arxiv.org/abs/2102.09725}{{\ttfamily
  2102.09725}}.

\bibitem{Lee:2021uqq}
R.~N. Lee, A.~von Manteuffel, R.~M. Schabinger, A.~V. Smirnov, V.~A. Smirnov
  and M.~Steinhauser, \emph{{Fermionic corrections to quark and gluon form
  factors in four-loop QCD}},
  \href{https://doi.org/10.1103/PhysRevD.104.074008}{\emph{Phys. Rev. D}
  {\bfseries 104} (2021) 074008}
  [\href{https://arxiv.org/abs/2105.11504}{{\ttfamily 2105.11504}}].

\bibitem{Lee:2022nhh}
R.~N. Lee, A.~von Manteuffel, R.~M. Schabinger, A.~V. Smirnov, V.~A. Smirnov
  and M.~Steinhauser, \emph{{Quark and gluon form factors in four-loop QCD}},
  \href{https://arxiv.org/abs/2202.04660}{{\ttfamily 2202.04660}}.

\bibitem{Kotikov:1990kg}
A.~Kotikov, \emph{{Differential equations method: New technique for massive
  Feynman diagrams calculation}},
  \href{https://doi.org/10.1016/0370-2693(91)90413-K}{\emph{Phys. Lett. B}
  {\bfseries 254} (1991) 158}.

\bibitem{Bern:1993kr}
Z.~Bern, L.~J. Dixon and D.~A. Kosower, \emph{{Dimensionally regulated pentagon
  integrals}}, \href{https://doi.org/10.1016/0550-3213(94)90398-0}{\emph{Nucl.
  Phys. B} {\bfseries 412} (1994) 751}
  [\href{https://arxiv.org/abs/hep-ph/9306240}{{\ttfamily hep-ph/9306240}}].

\bibitem{Remiddi:1997ny}
E.~Remiddi, \emph{{Differential equations for Feynman graph amplitudes}},
  \href{https://doi.org/10.1007/BF03185566}{\emph{Nuovo Cim. A} {\bfseries 110}
  (1997) 1435} [\href{https://arxiv.org/abs/hep-th/9711188}{{\ttfamily
  hep-th/9711188}}].

\bibitem{Henn:2013pwa}
J.~M. Henn, \emph{{Multiloop integrals in dimensional regularization made
  simple}}, \href{https://doi.org/10.1103/PhysRevLett.110.251601}{\emph{Phys.
  Rev. Lett.} {\bfseries 110} (2013) 251601}
  [\href{https://arxiv.org/abs/1304.1806}{{\ttfamily 1304.1806}}].

\bibitem{Brown:2008um}
F.~Brown, \emph{{The Massless higher-loop two-point function}},
  \href{https://doi.org/10.1007/s00220-009-0740-5}{\emph{Commun. Math. Phys.}
  {\bfseries 287} (2009) 925}
  [\href{https://arxiv.org/abs/0804.1660}{{\ttfamily 0804.1660}}].

\bibitem{Panzer:2014caa}
E.~Panzer, \emph{{Algorithms for the symbolic integration of hyperlogarithms
  with applications to Feynman integrals}},
  \href{https://doi.org/10.1016/j.cpc.2014.10.019}{\emph{Comput. Phys. Commun.}
  {\bfseries 188} (2015) 148}
  [\href{https://arxiv.org/abs/1403.3385}{{\ttfamily 1403.3385}}].

\bibitem{vonManteuffel:2015gxa}
A.~von Manteuffel, E.~Panzer and R.~M. Schabinger, \emph{{On the Computation of
  Form Factors in Massless QCD with Finite Master Integrals}},
  \href{https://doi.org/10.1103/PhysRevD.93.125014}{\emph{Phys. Rev. D}
  {\bfseries 93} (2016) 125014}
  [\href{https://arxiv.org/abs/1510.06758}{{\ttfamily 1510.06758}}].

\bibitem{Bonetti:2020hqh}
M.~Bonetti, E.~Panzer, V.~A. Smirnov and L.~Tancredi, \emph{{Two-loop mixed
  QCD-EW corrections to $gg \to Hg$}},
  \href{https://doi.org/10.1007/JHEP11(2020)045}{\emph{JHEP} {\bfseries 11}
  (2020) 045} [\href{https://arxiv.org/abs/2007.09813}{{\ttfamily
  2007.09813}}].

\bibitem{Remiddi:1999ew}
E.~Remiddi and J.~A.~M. Vermaseren, \emph{{Harmonic polylogarithms}},
  \href{https://doi.org/10.1142/S0217751X00000367}{\emph{Int. J. Mod. Phys. A}
  {\bfseries 15} (2000) 725}
  [\href{https://arxiv.org/abs/hep-ph/9905237}{{\ttfamily hep-ph/9905237}}].

\bibitem{Goncharov:2001iea}
A.~B. Goncharov, \emph{{Multiple polylogarithms and mixed Tate motives}},
  \href{https://arxiv.org/abs/math/0103059}{{\ttfamily math/0103059}}.

\bibitem{BrownLevin}
F.~Brown and A.~Levin, \emph{{Multiple Elliptic Polylogarithms}},
  \href{https://arxiv.org/abs/1110.6917}{{\ttfamily 1110.6917}}.

\bibitem{Bloch:2013tra}
S.~Bloch and P.~Vanhove, \emph{{The elliptic dilogarithm for the sunset
  graph}},  \href{https://arxiv.org/abs/1309.5865}{{\ttfamily 1309.5865}}.

\bibitem{Adams:2015gva}
L.~Adams, C.~Bogner and S.~Weinzierl, \emph{{The two-loop sunrise integral
  around four space-time dimensions and generalisations of the Clausen and
  Glaisher functions towards the elliptic case}},
  \href{https://arxiv.org/abs/1504.03255}{{\ttfamily 1504.03255}}.

\bibitem{Ablinger:2017bjx}
J.~Ablinger, J.~Bl\"umlein, A.~De~Freitas, M.~van Hoeij, E.~Imamoglu, C.~G.
  Raab et~al., \emph{{Iterated Elliptic and Hypergeometric Integrals for
  Feynman Diagrams}}, \href{https://doi.org/10.1063/1.4986417}{\emph{J. Math.
  Phys.} {\bfseries 59} (2018) 062305}
  [\href{https://arxiv.org/abs/1706.01299}{{\ttfamily 1706.01299}}].

\bibitem{Remiddi:2017har}
E.~Remiddi and L.~Tancredi, \emph{{An Elliptic Generalization of Multiple
  Polylogarithms}},
  \href{https://doi.org/10.1016/j.nuclphysb.2017.10.007}{\emph{Nucl. Phys. B}
  {\bfseries 925} (2017) 212}
  [\href{https://arxiv.org/abs/1709.03622}{{\ttfamily 1709.03622}}].

\bibitem{Broedel:2017kkb}
J.~Broedel, C.~Duhr, F.~Dulat and L.~Tancredi, \emph{{Elliptic polylogarithms
  and iterated integrals on elliptic curves. Part I: general formalism}},
  \href{https://doi.org/10.1007/JHEP05(2018)093}{\emph{JHEP} {\bfseries 05}
  (2018) 093} [\href{https://arxiv.org/abs/1712.07089}{{\ttfamily
  1712.07089}}].

\bibitem{Vollinga:2004sn}
J.~Vollinga and S.~Weinzierl, \emph{{Numerical evaluation of multiple
  polylogarithms}},
  \href{https://doi.org/10.1016/j.cpc.2004.12.009}{\emph{Comput. Phys. Commun.}
  {\bfseries 167} (2005) 177}
  [\href{https://arxiv.org/abs/hep-ph/0410259}{{\ttfamily hep-ph/0410259}}].

\bibitem{Naterop:2019xaf}
L.~Naterop, A.~Signer and Y.~Ulrich, \emph{{handyG \textemdash{}Rapid numerical
  evaluation of generalised polylogarithms in Fortran}},
  \href{https://doi.org/10.1016/j.cpc.2020.107165}{\emph{Comput. Phys. Commun.}
  {\bfseries 253} (2020) 107165}
  [\href{https://arxiv.org/abs/1909.01656}{{\ttfamily 1909.01656}}].

\bibitem{Duhr:2019rrs}
C.~Duhr and L.~Tancredi, \emph{{Algorithms and tools for iterated Eisenstein
  integrals}}, \href{https://doi.org/10.1007/JHEP02(2020)105}{\emph{JHEP}
  {\bfseries 02} (2020) 105}
  [\href{https://arxiv.org/abs/1912.00077}{{\ttfamily 1912.00077}}].

\bibitem{Walden:2020odh}
M.~Walden and S.~Weinzierl, \emph{{Numerical evaluation of iterated integrals
  related to elliptic Feynman integrals}},
  \href{https://doi.org/10.1016/j.cpc.2021.108020}{\emph{Comput. Phys. Commun.}
  {\bfseries 265} (2021) 108020}
  [\href{https://arxiv.org/abs/2010.05271}{{\ttfamily 2010.05271}}].

\bibitem{Bourjaily:2022bwx}
J.~L. Bourjaily et~al., \emph{{Functions Beyond Multiple Polylogarithms for
  Precision Collider Physics}},  3, 2022,
  \href{https://arxiv.org/abs/2203.07088}{{\ttfamily 2203.07088}}.

\bibitem{Binoth:2000ps}
T.~Binoth and G.~Heinrich, \emph{{An automatized algorithm to compute infrared
  divergent multiloop integrals}},
  \href{https://doi.org/10.1016/S0550-3213(00)00429-6}{\emph{Nucl. Phys. B}
  {\bfseries 585} (2000) 741}
  [\href{https://arxiv.org/abs/hep-ph/0004013}{{\ttfamily hep-ph/0004013}}].

\bibitem{Borowka:2015mxa}
S.~Borowka, G.~Heinrich, S.~P. Jones, M.~Kerner, J.~Schlenk and T.~Zirke,
  \emph{{SecDec-3.0: numerical evaluation of multi-scale integrals beyond one
  loop}}, \href{https://doi.org/10.1016/j.cpc.2015.05.022}{\emph{Comput. Phys.
  Commun.} {\bfseries 196} (2015) 470}
  [\href{https://arxiv.org/abs/1502.06595}{{\ttfamily 1502.06595}}].

\bibitem{Borowka:2018goh}
S.~Borowka, G.~Heinrich, S.~Jahn, S.~P. Jones, M.~Kerner and J.~Schlenk,
  \emph{{A GPU compatible quasi-Monte Carlo integrator interfaced to
  pySecDec}}, \href{https://doi.org/10.1016/j.cpc.2019.02.015}{\emph{Comput.
  Phys. Commun.} {\bfseries 240} (2019) 120}
  [\href{https://arxiv.org/abs/1811.11720}{{\ttfamily 1811.11720}}].

\bibitem{Smirnov:2021rhf}
A.~V. Smirnov, N.~D. Shapurov and L.~I. Vysotsky, \emph{{FIESTA5: numerical
  high-performance Feynman integral evaluation}},
  \href{https://arxiv.org/abs/2110.11660}{{\ttfamily 2110.11660}}.

\bibitem{Borowka:2016ypz}
S.~Borowka, N.~Greiner, G.~Heinrich, S.~P. Jones, M.~Kerner, J.~Schlenk et~al.,
  \emph{{Full top quark mass dependence in Higgs boson pair production at
  NLO}}, \href{https://doi.org/10.1007/JHEP10(2016)107}{\emph{JHEP} {\bfseries
  10} (2016) 107} [\href{https://arxiv.org/abs/1608.04798}{{\ttfamily
  1608.04798}}].

\bibitem{Jones:2018hbb}
S.~P. Jones, M.~Kerner and G.~Luisoni, \emph{{Next-to-Leading-Order QCD
  Corrections to Higgs Boson Plus Jet Production with Full Top-Quark Mass
  Dependence}},
  \href{https://doi.org/10.1103/PhysRevLett.120.162001}{\emph{Phys. Rev. Lett.}
  {\bfseries 120} (2018) 162001}
  [\href{https://arxiv.org/abs/1802.00349}{{\ttfamily 1802.00349}}].

\bibitem{Chen:2019fla}
L.~Chen, G.~Heinrich, S.~Jahn, S.~P. Jones, M.~Kerner, J.~Schlenk et~al.,
  \emph{{Photon pair production in gluon fusion: Top quark effects at NLO with
  threshold matching}},
  \href{https://doi.org/10.1007/JHEP04(2020)115}{\emph{JHEP} {\bfseries 04}
  (2020) 115} [\href{https://arxiv.org/abs/1911.09314}{{\ttfamily
  1911.09314}}].

\bibitem{Chen:2020gae}
L.~Chen, G.~Heinrich, S.~P. Jones, M.~Kerner, J.~Klappert and J.~Schlenk,
  \emph{{$ZH$ production in gluon fusion: two-loop amplitudes with full top
  quark mass dependence}},
  \href{https://doi.org/10.1007/JHEP03(2021)125}{\emph{JHEP} {\bfseries 03}
  (2021) 125} [\href{https://arxiv.org/abs/2011.12325}{{\ttfamily
  2011.12325}}].

\bibitem{vonManteuffel:2014qoa}
A.~von Manteuffel, E.~Panzer and R.~M. Schabinger, \emph{{A quasi-finite basis
  for multi-loop Feynman integrals}},
  \href{https://doi.org/10.1007/JHEP02(2015)120}{\emph{JHEP} {\bfseries 02}
  (2015) 120} [\href{https://arxiv.org/abs/1411.7392}{{\ttfamily 1411.7392}}].

\bibitem{vonManteuffel:2017myy}
A.~von Manteuffel and R.~M. Schabinger, \emph{{Numerical Multi-Loop
  Calculations via Finite Integrals and One-Mass EW-QCD Drell-Yan Master
  Integrals}}, \href{https://doi.org/10.1007/JHEP04(2017)129}{\emph{JHEP}
  {\bfseries 04} (2017) 129}
  [\href{https://arxiv.org/abs/1701.06583}{{\ttfamily 1701.06583}}].

\bibitem{Song:2021vru}
Q.~Song and A.~Freitas, \emph{{On the evaluation of two-loop electroweak box
  diagrams for $e^+e^- \to HZ$ production}},
  \href{https://doi.org/10.1007/JHEP04(2021)179}{\emph{JHEP} {\bfseries 04}
  (2021) 179} [\href{https://arxiv.org/abs/2101.00308}{{\ttfamily
  2101.00308}}].

\bibitem{Dubovyk:2018rlg}
I.~Dubovyk, A.~Freitas, J.~Gluza, T.~Riemann and J.~Usovitsch, \emph{{Complete
  electroweak two-loop corrections to Z boson production and decay}},
  \href{https://doi.org/10.1016/j.physletb.2018.06.037}{\emph{Phys. Lett. B}
  {\bfseries 783} (2018) 86}
  [\href{https://arxiv.org/abs/1804.10236}{{\ttfamily 1804.10236}}].

\bibitem{Pozzorini:2005ff}
S.~Pozzorini and E.~Remiddi, \emph{{Precise numerical evaluation of the two
  loop sunrise graph master integrals in the equal mass case}},
  \href{https://doi.org/10.1016/j.cpc.2006.05.005}{\emph{Comput. Phys. Commun.}
  {\bfseries 175} (2006) 381}
  [\href{https://arxiv.org/abs/hep-ph/0505041}{{\ttfamily hep-ph/0505041}}].

\bibitem{Aglietti:2007as}
U.~Aglietti, R.~Bonciani, L.~Grassi and E.~Remiddi, \emph{{The Two loop crossed
  ladder vertex diagram with two massive exchanges}},
  \href{https://doi.org/10.1016/j.nuclphysb.2007.07.019}{\emph{Nucl. Phys. B}
  {\bfseries 789} (2008) 45} [\href{https://arxiv.org/abs/0705.2616}{{\ttfamily
  0705.2616}}].

\bibitem{Lee:2017qql}
R.~N. Lee, A.~V. Smirnov and V.~A. Smirnov, \emph{{Solving differential
  equations for Feynman integrals by expansions near singular points}},
  \href{https://doi.org/10.1007/JHEP03(2018)008}{\emph{JHEP} {\bfseries 03}
  (2018) 008} [\href{https://arxiv.org/abs/1709.07525}{{\ttfamily
  1709.07525}}].

\bibitem{Frellesvig:2019byn}
H.~Frellesvig, M.~Hidding, L.~Maestri, F.~Moriello and G.~Salvatori, \emph{{The
  complete set of two-loop master integrals for Higgs + jet production in
  QCD}}, \href{https://doi.org/10.1007/JHEP06(2020)093}{\emph{JHEP} {\bfseries
  06} (2020) 093} [\href{https://arxiv.org/abs/1911.06308}{{\ttfamily
  1911.06308}}].

\bibitem{Heller:2021gun}
M.~Heller, \emph{{Planar two-loop integrals for $\mathbf{\mu e}$ scattering in
  QED with finite lepton masses}},
  \href{https://arxiv.org/abs/2105.08046}{{\ttfamily 2105.08046}}.

\bibitem{Hidding:2020ytt}
M.~Hidding, \emph{{DiffExp, a Mathematica package for computing Feynman
  integrals in terms of one-dimensional series expansions}},
  \href{https://doi.org/10.1016/j.cpc.2021.108125}{\emph{Comput. Phys. Commun.}
  {\bfseries 269} (2021) 108125}
  [\href{https://arxiv.org/abs/2006.05510}{{\ttfamily 2006.05510}}].

\bibitem{Dubovyk:2022frj}
I.~Dubovyk, A.~Freitas, J.~Gluza, K.~Grzanka, M.~Hidding and J.~Usovitsch,
  \emph{{Evaluation of multi-loop multi-scale Feynman integrals for precision
  physics}},  \href{https://arxiv.org/abs/2201.02576}{{\ttfamily 2201.02576}}.

\bibitem{Liu:2022mfb}
Z.-F. Liu and Y.-Q. Ma, \emph{{Feynman integrals are completely determined by
  linear algebra}},  \href{https://arxiv.org/abs/2201.11637}{{\ttfamily
  2201.11637}}.

\bibitem{Liu:2017jxz}
X.~Liu, Y.-Q. Ma and C.-Y. Wang, \emph{{A Systematic and Efficient Method to
  Compute Multi-loop Master Integrals}},
  \href{https://doi.org/10.1016/j.physletb.2018.02.026}{\emph{Phys. Lett. B}
  {\bfseries 779} (2018) 353}
  [\href{https://arxiv.org/abs/1711.09572}{{\ttfamily 1711.09572}}].

\bibitem{Liu:2020kpc}
X.~Liu, Y.-Q. Ma, W.~Tao and P.~Zhang, \emph{{Calculation of Feynman loop
  integration and phase-space integration via auxiliary mass flow}},
  \href{https://doi.org/10.1088/1674-1137/abc538}{\emph{Chin. Phys. C}
  {\bfseries 45} (2021) 013115}
  [\href{https://arxiv.org/abs/2009.07987}{{\ttfamily 2009.07987}}].

\bibitem{Liu:2021wks}
X.~Liu and Y.-Q. Ma, \emph{{Multiloop corrections for collider processes using
  auxiliary mass flow}},  \href{https://arxiv.org/abs/2107.01864}{{\ttfamily
  2107.01864}}.

\bibitem{Liu:2022chg}
X.~Liu and Y.-Q. Ma, \emph{{AMFlow: a Mathematica package for Feynman integrals
  computation via Auxiliary Mass Flow}},
  \href{https://arxiv.org/abs/2201.11669}{{\ttfamily 2201.11669}}.

\bibitem{Bevilacqua:2011xh}
G.~Bevilacqua, M.~Czakon, M.~V. Garzelli, A.~van Hameren, A.~Kardos, C.~G.
  Papadopoulos et~al., \emph{{HELAC-NLO}},
  \href{https://doi.org/10.1016/j.cpc.2012.10.033}{\emph{Comput. Phys. Commun.}
  {\bfseries 184} (2013) 986}
  [\href{https://arxiv.org/abs/1110.1499}{{\ttfamily 1110.1499}}].

\bibitem{Alwall:2014hca}
J.~Alwall, R.~Frederix, S.~Frixione, V.~Hirschi, F.~Maltoni, O.~Mattelaer
  et~al., \emph{{The automated computation of tree-level and next-to-leading
  order differential cross sections, and their matching to parton shower
  simulations}}, \href{https://doi.org/10.1007/JHEP07(2014)079}{\emph{JHEP}
  {\bfseries 07} (2014) 079} [\href{https://arxiv.org/abs/1405.0301}{{\ttfamily
  1405.0301}}].

\bibitem{Frederix:2018nkq}
R.~Frederix, S.~Frixione, V.~Hirschi, D.~Pagani, H.~S. Shao and M.~Zaro,
  \emph{{The automation of next-to-leading order electroweak calculations}},
  \href{https://doi.org/10.1007/JHEP11(2021)085}{\emph{JHEP} {\bfseries 07}
  (2018) 185} [\href{https://arxiv.org/abs/1804.10017}{{\ttfamily
  1804.10017}}].

\bibitem{Figueroa:2021txg}
D.~Figueroa, S.~Quackenbush, L.~Reina and C.~Reuschle, \emph{{Updates to the
  one-loop provider NLOX}},
  \href{https://doi.org/10.1016/j.cpc.2021.108150}{\emph{Comput. Phys. Commun.}
  {\bfseries 270} (2022) 108150}
  [\href{https://arxiv.org/abs/2101.01305}{{\ttfamily 2101.01305}}].

\bibitem{Buccioni:2019sur}
F.~Buccioni, J.-N. Lang, J.~M. Lindert, P.~Maierh\"ofer, S.~Pozzorini, H.~Zhang
  et~al., \emph{{OpenLoops 2}},
  \href{https://doi.org/10.1140/epjc/s10052-019-7306-2}{\emph{Eur. Phys. J. C}
  {\bfseries 79} (2019) 866}
  [\href{https://arxiv.org/abs/1907.13071}{{\ttfamily 1907.13071}}].

\bibitem{Denner:2017wsf}
A.~Denner, J.-N. Lang and S.~Uccirati, \emph{{Recola2: REcursive Computation of
  One-Loop Amplitudes 2}},
  \href{https://doi.org/10.1016/j.cpc.2017.11.013}{\emph{Comput. Phys. Commun.}
  {\bfseries 224} (2018) 346}
  [\href{https://arxiv.org/abs/1711.07388}{{\ttfamily 1711.07388}}].

\bibitem{Borowka:2016ehy}
S.~Borowka, N.~Greiner, G.~Heinrich, S.~P. Jones, M.~Kerner, J.~Schlenk et~al.,
  \emph{{Higgs Boson Pair Production in Gluon Fusion at Next-to-Leading Order
  with Full Top-Quark Mass Dependence}},
  \href{https://doi.org/10.1103/PhysRevLett.117.079901}{\emph{Phys. Rev. Lett.}
  {\bfseries 117} (2016) 012001}
  [\href{https://arxiv.org/abs/1604.06447}{{\ttfamily 1604.06447}}].

\bibitem{Heller:2020owb}
M.~Heller, A.~von Manteuffel, R.~M. Schabinger and H.~Spiesberger, \emph{{Mixed
  EW-QCD two-loop amplitudes for $q\bar{q} \to \ell^+\ell^-$ and $\gamma_5$
  scheme independence of multi-loop corrections}},
  \href{https://doi.org/10.1007/JHEP05(2021)213}{\emph{JHEP} {\bfseries 05}
  (2021) 213} [\href{https://arxiv.org/abs/2012.05918}{{\ttfamily
  2012.05918}}].

\bibitem{Bonciani:2021zzf}
R.~Bonciani, L.~Buonocore, M.~Grazzini, S.~Kallweit, N.~Rana, F.~Tramontano
  et~al., \emph{{Mixed Strong-Electroweak Corrections to the Drell-Yan
  Process}}, \href{https://doi.org/10.1103/PhysRevLett.128.012002}{\emph{Phys.
  Rev. Lett.} {\bfseries 128} (2022) 012002}
  [\href{https://arxiv.org/abs/2106.11953}{{\ttfamily 2106.11953}}].

\bibitem{Becchetti:2021axs}
M.~Becchetti, F.~Moriello and A.~Schweitzer, \emph{{Two-loop amplitude for
  mixed QCD-EW corrections to $gg \to Hg$}},
  \href{https://arxiv.org/abs/2112.07578}{{\ttfamily 2112.07578}}.

\bibitem{Badger:2018enw}
S.~Badger, C.~Br\o{}nnum-Hansen, H.~B. Hartanto and T.~Peraro, \emph{{Analytic
  helicity amplitudes for two-loop five-gluon scattering: the single-minus
  case}}, \href{https://doi.org/10.1007/JHEP01(2019)186}{\emph{JHEP} {\bfseries
  01} (2019) 186} [\href{https://arxiv.org/abs/1811.11699}{{\ttfamily
  1811.11699}}].

\bibitem{Abreu:2018zmy}
S.~Abreu, J.~Dormans, F.~Febres~Cordero, H.~Ita and B.~Page, \emph{{Analytic
  Form of Planar Two-Loop Five-Gluon Scattering Amplitudes in QCD}},
  \href{https://doi.org/10.1103/PhysRevLett.122.082002}{\emph{Phys. Rev. Lett.}
  {\bfseries 122} (2019) 082002}
  [\href{https://arxiv.org/abs/1812.04586}{{\ttfamily 1812.04586}}].

\bibitem{Abreu:2019odu}
S.~Abreu, J.~Dormans, F.~Febres~Cordero, H.~Ita, B.~Page and V.~Sotnikov,
  \emph{{Analytic Form of the Planar Two-Loop Five-Parton Scattering Amplitudes
  in QCD}}, \href{https://doi.org/10.1007/JHEP05(2019)084}{\emph{JHEP}
  {\bfseries 05} (2019) 084}
  [\href{https://arxiv.org/abs/1904.00945}{{\ttfamily 1904.00945}}].

\bibitem{Badger:2019djh}
S.~Badger, D.~Chicherin, T.~Gehrmann, G.~Heinrich, J.~M. Henn, T.~Peraro
  et~al., \emph{{Analytic form of the full two-loop five-gluon all-plus
  helicity amplitude}},
  \href{https://doi.org/10.1103/PhysRevLett.123.071601}{\emph{Phys. Rev. Lett.}
  {\bfseries 123} (2019) 071601}
  [\href{https://arxiv.org/abs/1905.03733}{{\ttfamily 1905.03733}}].

\bibitem{Abreu:2020cwb}
S.~Abreu, B.~Page, E.~Pascual and V.~Sotnikov, \emph{{Leading-Color Two-Loop
  QCD Corrections for Three-Photon Production at Hadron Colliders}},
  \href{https://doi.org/10.1007/JHEP01(2021)078}{\emph{JHEP} {\bfseries 01}
  (2021) 078} [\href{https://arxiv.org/abs/2010.15834}{{\ttfamily
  2010.15834}}].

\bibitem{Chawdhry:2020for}
H.~A. Chawdhry, M.~Czakon, A.~Mitov and R.~Poncelet, \emph{{Two-loop
  leading-color helicity amplitudes for three-photon production at the LHC}},
  \href{https://doi.org/10.1007/JHEP06(2021)150}{\emph{JHEP} {\bfseries 06}
  (2021) 150} [\href{https://arxiv.org/abs/2012.13553}{{\ttfamily
  2012.13553}}].

\bibitem{Agarwal:2021grm}
B.~Agarwal, F.~Buccioni, A.~von Manteuffel and L.~Tancredi, \emph{{Two-loop
  leading colour QCD corrections to $q \bar{q} \to \gamma \gamma g$ and $q g
  \to \gamma \gamma q$}},
  \href{https://doi.org/10.1007/JHEP04(2021)201}{\emph{JHEP} {\bfseries 04}
  (2021) 201} [\href{https://arxiv.org/abs/2102.01820}{{\ttfamily
  2102.01820}}].

\bibitem{Abreu:2021oya}
S.~Abreu, F.~Febres~Cordero, H.~Ita, B.~Page and V.~Sotnikov,
  \emph{{Leading-color two-loop QCD corrections for three-jet production at
  hadron colliders}},
  \href{https://doi.org/10.1007/JHEP07(2021)095}{\emph{JHEP} {\bfseries 07}
  (2021) 095} [\href{https://arxiv.org/abs/2102.13609}{{\ttfamily
  2102.13609}}].

\bibitem{Chawdhry:2021mkw}
H.~A. Chawdhry, M.~Czakon, A.~Mitov and R.~Poncelet, \emph{{Two-loop
  leading-colour QCD helicity amplitudes for two-photon plus jet production at
  the LHC}}, \href{https://doi.org/10.1007/JHEP07(2021)164}{\emph{JHEP}
  {\bfseries 07} (2021) 164}
  [\href{https://arxiv.org/abs/2103.04319}{{\ttfamily 2103.04319}}].

\bibitem{Agarwal:2021vdh}
B.~Agarwal, F.~Buccioni, A.~von Manteuffel and L.~Tancredi, \emph{{Two-Loop
  Helicity Amplitudes for Diphoton Plus Jet Production in Full Color}},
  \href{https://doi.org/10.1103/PhysRevLett.127.262001}{\emph{Phys. Rev. Lett.}
  {\bfseries 127} (2021) 262001}
  [\href{https://arxiv.org/abs/2105.04585}{{\ttfamily 2105.04585}}].

\bibitem{Badger:2021ega}
S.~Badger, H.~B. Hartanto, J.~Kry\'s and S.~Zoia, \emph{{Two-loop
  leading-colour QCD helicity amplitudes for Higgs boson production in
  association with a bottom-quark pair at the LHC}},
  \href{https://doi.org/10.1007/JHEP11(2021)012}{\emph{JHEP} {\bfseries 11}
  (2021) 012} [\href{https://arxiv.org/abs/2107.14733}{{\ttfamily
  2107.14733}}].

\bibitem{Abreu:2021asb}
S.~Abreu, F.~Febres~Cordero, H.~Ita, M.~Klinkert, B.~Page and V.~Sotnikov,
  \emph{{Leading-Color Two-Loop Amplitudes for Four Partons and a W Boson in
  QCD}},  \href{https://arxiv.org/abs/2110.07541}{{\ttfamily 2110.07541}}.

\bibitem{Badger:2022ncb}
S.~Badger, H.~B. Hartanto, J.~Kry\'s and S.~Zoia, \emph{{Two-loop leading
  colour helicity amplitudes for $W^\pm\gamma+j$ production at the LHC}},
  \href{https://arxiv.org/abs/2201.04075}{{\ttfamily 2201.04075}}.

\bibitem{Henn:2016tyf}
J.~Henn, A.~V. Smirnov, V.~A. Smirnov and M.~Steinhauser, \emph{{Massive
  three-loop form factor in the planar limit}},
  \href{https://doi.org/10.1007/JHEP01(2017)074}{\emph{JHEP} {\bfseries 01}
  (2017) 074} [\href{https://arxiv.org/abs/1611.07535}{{\ttfamily
  1611.07535}}].

\bibitem{Lee:2018nxa}
R.~N. Lee, A.~V. Smirnov, V.~A. Smirnov and M.~Steinhauser, \emph{{Three-loop
  massive form factors: complete light-fermion corrections for the vector
  current}}, \href{https://doi.org/10.1007/JHEP03(2018)136}{\emph{JHEP}
  {\bfseries 03} (2018) 136}
  [\href{https://arxiv.org/abs/1801.08151}{{\ttfamily 1801.08151}}].

\bibitem{Ablinger:2018yae}
J.~Ablinger, J.~Bl\"umlein, P.~Marquard, N.~Rana and C.~Schneider, \emph{{Heavy
  quark form factors at three loops in the planar limit}},
  \href{https://doi.org/10.1016/j.physletb.2018.05.077}{\emph{Phys. Lett. B}
  {\bfseries 782} (2018) 528}
  [\href{https://arxiv.org/abs/1804.07313}{{\ttfamily 1804.07313}}].

\bibitem{Blumlein:2019oas}
J.~Bl\"umlein, P.~Marquard, N.~Rana and C.~Schneider, \emph{{The Heavy Fermion
  Contributions to the Massive Three Loop Form Factors}},
  \href{https://doi.org/10.1016/j.nuclphysb.2019.114751}{\emph{Nucl. Phys. B}
  {\bfseries 949} (2019) 114751}
  [\href{https://arxiv.org/abs/1908.00357}{{\ttfamily 1908.00357}}].

\bibitem{Czakon:2021yub}
M.~Czakon, R.~V. Harlander, J.~Klappert and M.~Niggetiedt, \emph{{Exact
  Top-Quark Mass Dependence in Hadronic Higgs Production}},
  \href{https://doi.org/10.1103/PhysRevLett.127.162002}{\emph{Phys. Rev. Lett.}
  {\bfseries 127} (2021) 162002}
  [\href{https://arxiv.org/abs/2105.04436}{{\ttfamily 2105.04436}}].

\bibitem{Fael:2022rgm}
M.~Fael, F.~Lange, K.~Sch\"onwald and M.~Steinhauser, \emph{{Massive vector
  form factors to three loops}},
  \href{https://arxiv.org/abs/2202.05276}{{\ttfamily 2202.05276}}.

\bibitem{Caola:2020dfu}
F.~Caola, A.~von Manteuffel and L.~Tancredi, \emph{{Diphoton Amplitudes in
  Three-Loop Quantum Chromodynamics}},
  \href{https://doi.org/10.1103/PhysRevLett.126.112004}{\emph{Phys. Rev. Lett.}
  {\bfseries 126} (2021) 112004}
  [\href{https://arxiv.org/abs/2011.13946}{{\ttfamily 2011.13946}}].

\bibitem{Caola:2021rqz}
F.~Caola, A.~Chakraborty, G.~Gambuti, A.~von Manteuffel and L.~Tancredi,
  \emph{{Three-loop helicity amplitudes for four-quark scattering in massless
  QCD}}, \href{https://doi.org/10.1007/JHEP10(2021)206}{\emph{JHEP} {\bfseries
  10} (2021) 206} [\href{https://arxiv.org/abs/2108.00055}{{\ttfamily
  2108.00055}}].

\bibitem{Bargiela:2021wuy}
P.~Bargiela, F.~Caola, A.~von Manteuffel and L.~Tancredi, \emph{{Three-loop
  helicity amplitudes for diphoton production in gluon fusion}},
  \href{https://doi.org/10.1007/JHEP02(2022)153}{\emph{JHEP} {\bfseries 02}
  (2022) 153} [\href{https://arxiv.org/abs/2111.13595}{{\ttfamily
  2111.13595}}].

\bibitem{Caola:2021izf}
F.~Caola, A.~Chakraborty, G.~Gambuti, A.~von Manteuffel and L.~Tancredi,
  \emph{{Three-loop gluon scattering in QCD and the gluon Regge trajectory}},
  \href{https://arxiv.org/abs/2112.11097}{{\ttfamily 2112.11097}}.

\bibitem{Chakraborty:2022yan}
A.~Chakraborty, T.~Huber, R.~N. Lee, A.~von Manteuffel, R.~M. Schabinger, A.~V.
  Smirnov et~al., \emph{{The $Hb\bar{b}$ vertex at four loops and hard matching
  coefficients in SCET for various currents}},
  \href{https://arxiv.org/abs/2204.02422}{{\ttfamily 2204.02422}}.

\bibitem{Herzog:2017ohr}
F.~Herzog, B.~Ruijl, T.~Ueda, J.~A.~M. Vermaseren and A.~Vogt, \emph{{The
  five-loop beta function of Yang-Mills theory with fermions}},
  \href{https://doi.org/10.1007/JHEP02(2017)090}{\emph{JHEP} {\bfseries 02}
  (2017) 090} [\href{https://arxiv.org/abs/1701.01404}{{\ttfamily
  1701.01404}}].

\bibitem{Luthe:2017ttg}
T.~Luthe, A.~Maier, P.~Marquard and Y.~Schroder, \emph{{The five-loop Beta
  function for a general gauge group and anomalous dimensions beyond Feynman
  gauge}}, \href{https://doi.org/10.1007/JHEP10(2017)166}{\emph{JHEP}
  {\bfseries 10} (2017) 166}
  [\href{https://arxiv.org/abs/1709.07718}{{\ttfamily 1709.07718}}].

\bibitem{Moch:2017uml}
S.~Moch, B.~Ruijl, T.~Ueda, J.~A.~M. Vermaseren and A.~Vogt, \emph{{Four-Loop
  Non-Singlet Splitting Functions in the Planar Limit and Beyond}},
  \href{https://doi.org/10.1007/JHEP10(2017)041}{\emph{JHEP} {\bfseries 10}
  (2017) 041} [\href{https://arxiv.org/abs/1707.08315}{{\ttfamily
  1707.08315}}].

\bibitem{Moch:2021qrk}
S.~Moch, B.~Ruijl, T.~Ueda, J.~A.~M. Vermaseren and A.~Vogt, \emph{{Low moments
  of the four-loop splitting functions in QCD}},
  \href{https://doi.org/10.1016/j.physletb.2021.136853}{\emph{Phys. Lett. B}
  {\bfseries 825} (2022) 136853}
  [\href{https://arxiv.org/abs/2111.15561}{{\ttfamily 2111.15561}}].

\bibitem{Chetyrkin:1981qh}
K.~G. Chetyrkin and F.~V. Tkachov, \emph{{Integration by Parts: The Algorithm
  to Calculate beta Functions in 4 Loops}},
  \href{https://doi.org/10.1016/0550-3213(81)90199-1}{\emph{Nucl. Phys. B}
  {\bfseries 192} (1981) 159}.

\bibitem{Peraro:2019cjj}
T.~Peraro and L.~Tancredi, \emph{{Physical projectors for multi-leg helicity
  amplitudes}}, \href{https://doi.org/10.1007/JHEP07(2019)114}{\emph{JHEP}
  {\bfseries 07} (2019) 114}
  [\href{https://arxiv.org/abs/1906.03298}{{\ttfamily 1906.03298}}].

\bibitem{Peraro:2020sfm}
T.~Peraro and L.~Tancredi, \emph{{Tensor decomposition for bosonic and
  fermionic scattering amplitudes}},
  \href{https://doi.org/10.1103/PhysRevD.103.054042}{\emph{Phys. Rev. D}
  {\bfseries 103} (2021) 054042}
  [\href{https://arxiv.org/abs/2012.00820}{{\ttfamily 2012.00820}}].

\bibitem{Chen:2019wyb}
L.~Chen, \emph{{A prescription for projectors to compute helicity amplitudes in
  D dimensions}},
  \href{https://doi.org/10.1140/epjc/s10052-021-09210-9}{\emph{Eur. Phys. J. C}
  {\bfseries 81} (2021) 417}
  [\href{https://arxiv.org/abs/1904.00705}{{\ttfamily 1904.00705}}].

\bibitem{Mastrolia:2011pr}
P.~Mastrolia and G.~Ossola, \emph{{On the Integrand-Reduction Method for
  Two-Loop Scattering Amplitudes}},
  \href{https://doi.org/10.1007/JHEP11(2011)014}{\emph{JHEP} {\bfseries 11}
  (2011) 014} [\href{https://arxiv.org/abs/1107.6041}{{\ttfamily 1107.6041}}].

\bibitem{Badger:2012dp}
S.~Badger, H.~Frellesvig and Y.~Zhang, \emph{{Hepta-Cuts of Two-Loop Scattering
  Amplitudes}}, \href{https://doi.org/10.1007/JHEP04(2012)055}{\emph{JHEP}
  {\bfseries 04} (2012) 055} [\href{https://arxiv.org/abs/1202.2019}{{\ttfamily
  1202.2019}}].

\bibitem{Zhang:2012ce}
Y.~Zhang, \emph{{Integrand-Level Reduction of Loop Amplitudes by Computational
  Algebraic Geometry Methods}},
  \href{https://doi.org/10.1007/JHEP09(2012)042}{\emph{JHEP} {\bfseries 09}
  (2012) 042} [\href{https://arxiv.org/abs/1205.5707}{{\ttfamily 1205.5707}}].

\bibitem{Mastrolia:2016dhn}
P.~Mastrolia, T.~Peraro and A.~Primo, \emph{{Adaptive Integrand Decomposition
  in parallel and orthogonal space}},
  \href{https://doi.org/10.1007/JHEP08(2016)164}{\emph{JHEP} {\bfseries 08}
  (2016) 164} [\href{https://arxiv.org/abs/1605.03157}{{\ttfamily
  1605.03157}}].

\bibitem{delAguila:2004nf}
F.~del Aguila and R.~Pittau, \emph{{Recursive numerical calculus of one-loop
  tensor integrals}},
  \href{https://doi.org/10.1088/1126-6708/2004/07/017}{\emph{JHEP} {\bfseries
  07} (2004) 017} [\href{https://arxiv.org/abs/hep-ph/0404120}{{\ttfamily
  hep-ph/0404120}}].

\bibitem{Ossola:2006us}
G.~Ossola, C.~G. Papadopoulos and R.~Pittau, \emph{{Reducing full one-loop
  amplitudes to scalar integrals at the integrand level}},
  \href{https://doi.org/10.1016/j.nuclphysb.2006.11.012}{\emph{Nucl. Phys. B}
  {\bfseries 763} (2007) 147}
  [\href{https://arxiv.org/abs/hep-ph/0609007}{{\ttfamily hep-ph/0609007}}].

\bibitem{Ossola:2007bb}
G.~Ossola, C.~G. Papadopoulos and R.~Pittau, \emph{{Numerical evaluation of
  six-photon amplitudes}},
  \href{https://doi.org/10.1088/1126-6708/2007/07/085}{\emph{JHEP} {\bfseries
  07} (2007) 085} [\href{https://arxiv.org/abs/0704.1271}{{\ttfamily
  0704.1271}}].

\bibitem{Ita:2015tya}
H.~Ita, \emph{{Two-loop Integrand Decomposition into Master Integrals and
  Surface Terms}},
  \href{https://doi.org/10.1103/PhysRevD.94.116015}{\emph{Phys. Rev. D}
  {\bfseries 94} (2016) 116015}
  [\href{https://arxiv.org/abs/1510.05626}{{\ttfamily 1510.05626}}].

\bibitem{Ellis:2007br}
R.~K. Ellis, W.~T. Giele and Z.~Kunszt, \emph{{A Numerical Unitarity Formalism
  for Evaluating One-Loop Amplitudes}},
  \href{https://doi.org/10.1088/1126-6708/2008/03/003}{\emph{JHEP} {\bfseries
  03} (2008) 003} [\href{https://arxiv.org/abs/0708.2398}{{\ttfamily
  0708.2398}}].

\bibitem{Giele:2008ve}
W.~T. Giele, Z.~Kunszt and K.~Melnikov, \emph{{Full one-loop amplitudes from
  tree amplitudes}},
  \href{https://doi.org/10.1088/1126-6708/2008/04/049}{\emph{JHEP} {\bfseries
  04} (2008) 049} [\href{https://arxiv.org/abs/0801.2237}{{\ttfamily
  0801.2237}}].

\bibitem{Berger:2008sj}
C.~F. Berger, Z.~Bern, L.~J. Dixon, F.~Febres~Cordero, D.~Forde, H.~Ita et~al.,
  \emph{{An Automated Implementation of On-Shell Methods for One-Loop
  Amplitudes}}, \href{https://doi.org/10.1103/PhysRevD.78.036003}{\emph{Phys.
  Rev. D} {\bfseries 78} (2008) 036003}
  [\href{https://arxiv.org/abs/0803.4180}{{\ttfamily 0803.4180}}].

\bibitem{Abreu:2017xsl}
S.~Abreu, F.~Febres~Cordero, H.~Ita, M.~Jaquier, B.~Page and M.~Zeng,
  \emph{{Two-Loop Four-Gluon Amplitudes from Numerical Unitarity}},
  \href{https://doi.org/10.1103/PhysRevLett.119.142001}{\emph{Phys. Rev. Lett.}
  {\bfseries 119} (2017) 142001}
  [\href{https://arxiv.org/abs/1703.05273}{{\ttfamily 1703.05273}}].

\bibitem{Abreu:2017idw}
S.~Abreu, F.~Febres~Cordero, H.~Ita, M.~Jaquier and B.~Page, \emph{{Subleading
  Poles in the Numerical Unitarity Method at Two Loops}},
  \href{https://doi.org/10.1103/PhysRevD.95.096011}{\emph{Phys. Rev. D}
  {\bfseries 95} (2017) 096011}
  [\href{https://arxiv.org/abs/1703.05255}{{\ttfamily 1703.05255}}].

\bibitem{Abreu:2017hqn}
S.~Abreu, F.~Febres~Cordero, H.~Ita, B.~Page and M.~Zeng, \emph{{Planar
  Two-Loop Five-Gluon Amplitudes from Numerical Unitarity}},
  \href{https://doi.org/10.1103/PhysRevD.97.116014}{\emph{Phys. Rev. D}
  {\bfseries 97} (2018) 116014}
  [\href{https://arxiv.org/abs/1712.03946}{{\ttfamily 1712.03946}}].

\bibitem{Abreu:2018jgq}
S.~Abreu, F.~Febres~Cordero, H.~Ita, B.~Page and V.~Sotnikov, \emph{{Planar
  Two-Loop Five-Parton Amplitudes from Numerical Unitarity}},
  \href{https://doi.org/10.1007/JHEP11(2018)116}{\emph{JHEP} {\bfseries 11}
  (2018) 116} [\href{https://arxiv.org/abs/1809.09067}{{\ttfamily
  1809.09067}}].

\bibitem{Pozzorini:2020hkx}
S.~Pozzorini, H.~Zhang and M.~F. Zoller, \emph{{Rational Terms of UV Origin at
  Two Loops}}, \href{https://doi.org/10.1007/JHEP05(2020)077}{\emph{JHEP}
  {\bfseries 05} (2020) 077}
  [\href{https://arxiv.org/abs/2001.11388}{{\ttfamily 2001.11388}}].

\bibitem{Lang:2020nnl}
J.-N. Lang, S.~Pozzorini, H.~Zhang and M.~F. Zoller, \emph{{Two-Loop Rational
  Terms in Yang-Mills Theories}},
  \href{https://doi.org/10.1007/JHEP10(2020)016}{\emph{JHEP} {\bfseries 10}
  (2020) 016} [\href{https://arxiv.org/abs/2007.03713}{{\ttfamily
  2007.03713}}].

\bibitem{Lang:2021hnw}
J.-N. Lang, S.~Pozzorini, H.~Zhang and M.~F. Zoller, \emph{{Two-loop rational
  terms for spontaneously broken theories}},
  \href{https://doi.org/10.1007/JHEP01(2022)105}{\emph{JHEP} {\bfseries 01}
  (2022) 105} [\href{https://arxiv.org/abs/2107.10288}{{\ttfamily
  2107.10288}}].

\bibitem{Pozzorini:2022ohr}
S.~Pozzorini, N.~Sch\"ar and M.~F. Zoller, \emph{{Two-loop tensor integral
  coefficients in OpenLoops}},
  \href{https://arxiv.org/abs/2201.11615}{{\ttfamily 2201.11615}}.

\bibitem{Laporta:2000dsw}
S.~Laporta, \emph{{High precision calculation of multiloop Feynman integrals by
  difference equations}},
  \href{https://doi.org/10.1142/S0217751X00002159}{\emph{Int. J. Mod. Phys. A}
  {\bfseries 15} (2000) 5087}
  [\href{https://arxiv.org/abs/hep-ph/0102033}{{\ttfamily hep-ph/0102033}}].

\bibitem{Smirnov:2019qkx}
A.~V. Smirnov and F.~S. Chuharev, \emph{{FIRE6: Feynman Integral REduction with
  Modular Arithmetic}},
  \href{https://doi.org/10.1016/j.cpc.2019.106877}{\emph{Comput. Phys. Commun.}
  {\bfseries 247} (2020) 106877}
  [\href{https://arxiv.org/abs/1901.07808}{{\ttfamily 1901.07808}}].

\bibitem{vonManteuffel:2012np}
A.~von Manteuffel and C.~Studerus, \emph{{{\tt Reduze 2} - Distributed Feynman
  Integral Reduction}},  \href{https://arxiv.org/abs/1201.4330}{{\ttfamily
  1201.4330}}.

\bibitem{Lee:2013mka}
R.~N. Lee, \emph{{LiteRed 1.4: a powerful tool for reduction of multiloop
  integrals}}, \href{https://doi.org/10.1088/1742-6596/523/1/012059}{\emph{J.
  Phys. Conf. Ser.} {\bfseries 523} (2014) 012059}
  [\href{https://arxiv.org/abs/1310.1145}{{\ttfamily 1310.1145}}].

\bibitem{Klappert:2020nbg}
J.~Klappert, F.~Lange, P.~Maierh\"ofer and J.~Usovitsch, \emph{{Integral
  reduction with Kira 2.0 and finite field methods}},
  \href{https://doi.org/10.1016/j.cpc.2021.108024}{\emph{Comput. Phys. Commun.}
  {\bfseries 266} (2021) 108024}
  [\href{https://arxiv.org/abs/2008.06494}{{\ttfamily 2008.06494}}].

\bibitem{Gluza:2010ws}
J.~Gluza, K.~Kajda and D.~A. Kosower, \emph{{Towards a Basis for Planar
  Two-Loop Integrals}},
  \href{https://doi.org/10.1103/PhysRevD.83.045012}{\emph{Phys. Rev. D}
  {\bfseries 83} (2011) 045012}
  [\href{https://arxiv.org/abs/1009.0472}{{\ttfamily 1009.0472}}].

\bibitem{Larsen:2015ped}
K.~J. Larsen and Y.~Zhang, \emph{{Integration-by-parts reductions from
  unitarity cuts and algebraic geometry}},
  \href{https://doi.org/10.1103/PhysRevD.93.041701}{\emph{Phys. Rev. D}
  {\bfseries 93} (2016) 041701}
  [\href{https://arxiv.org/abs/1511.01071}{{\ttfamily 1511.01071}}].

\bibitem{Georgoudis:2016wff}
A.~Georgoudis, K.~J. Larsen and Y.~Zhang, \emph{{Azurite: An algebraic geometry
  based package for finding bases of loop integrals}},
  \href{https://doi.org/10.1016/j.cpc.2017.08.013}{\emph{Comput. Phys. Commun.}
  {\bfseries 221} (2017) 203}
  [\href{https://arxiv.org/abs/1612.04252}{{\ttfamily 1612.04252}}].

\bibitem{Bohm:2017qme}
J.~B\"ohm, A.~Georgoudis, K.~J. Larsen, M.~Schulze and Y.~Zhang,
  \emph{{Complete sets of logarithmic vector fields for integration-by-parts
  identities of Feynman integrals}},
  \href{https://doi.org/10.1103/PhysRevD.98.025023}{\emph{Phys. Rev. D}
  {\bfseries 98} (2018) 025023}
  [\href{https://arxiv.org/abs/1712.09737}{{\ttfamily 1712.09737}}].

\bibitem{Lee:2014tja}
R.~N. Lee, \emph{{Modern techniques of multiloop calculations}},  in
  \emph{{49th Rencontres de Moriond on QCD and High Energy Interactions}},
  pp.~297--300, 2014, \href{https://arxiv.org/abs/1405.5616}{{\ttfamily
  1405.5616}}.

\bibitem{Bitoun:2017nre}
T.~Bitoun, C.~Bogner, R.~P. Klausen and E.~Panzer, \emph{{Feynman integral
  relations from parametric annihilators}},
  \href{https://doi.org/10.1007/s11005-018-1114-8}{\emph{Lett. Math. Phys.}
  {\bfseries 109} (2019) 497}
  [\href{https://arxiv.org/abs/1712.09215}{{\ttfamily 1712.09215}}].

\bibitem{Smirnov:2020quc}
A.~Smirnov and V.~Smirnov, \emph{{How to choose master integrals}},
  \href{https://arxiv.org/abs/2002.08042}{{\ttfamily 2002.08042}}.

\bibitem{Usovitsch:2020jrk}
J.~Usovitsch, \emph{{Factorization of denominators in integration-by-parts
  reductions}},  \href{https://arxiv.org/abs/2002.08173}{{\ttfamily
  2002.08173}}.

\bibitem{Mastrolia:2018uzb}
P.~Mastrolia and S.~Mizera, \emph{{Feynman Integrals and Intersection Theory}},
  \href{https://doi.org/10.1007/JHEP02(2019)139}{\emph{JHEP} {\bfseries 02}
  (2019) 139} [\href{https://arxiv.org/abs/1810.03818}{{\ttfamily
  1810.03818}}].

\bibitem{Frellesvig:2019kgj}
H.~Frellesvig, F.~Gasparotto, S.~Laporta, M.~K. Mandal, P.~Mastrolia,
  L.~Mattiazzi et~al., \emph{{Decomposition of Feynman Integrals on the Maximal
  Cut by Intersection Numbers}},
  \href{https://doi.org/10.1007/JHEP05(2019)153}{\emph{JHEP} {\bfseries 05}
  (2019) 153} [\href{https://arxiv.org/abs/1901.11510}{{\ttfamily
  1901.11510}}].

\bibitem{vonManteuffel:2014ixa}
A.~von Manteuffel and R.~M. Schabinger, \emph{{A novel approach to integration
  by parts reduction}},
  \href{https://doi.org/10.1016/j.physletb.2015.03.029}{\emph{Phys. Lett.}
  {\bfseries B744} (2015) 101}
  [\href{https://arxiv.org/abs/1406.4513}{{\ttfamily 1406.4513}}].

\bibitem{Peraro:2016wsq}
T.~Peraro, \emph{{Scattering amplitudes over finite fields and multivariate
  functional reconstruction}},
  \href{https://doi.org/10.1007/JHEP12(2016)030}{\emph{JHEP} {\bfseries 12}
  (2016) 030} [\href{https://arxiv.org/abs/1608.01902}{{\ttfamily
  1608.01902}}].

\bibitem{Klappert:2019emp}
J.~Klappert and F.~Lange, \emph{{Reconstructing rational functions with
  FireFly}}, \href{https://doi.org/10.1016/j.cpc.2019.106951}{\emph{Comput.
  Phys. Commun.} {\bfseries 247} (2020) 106951}
  [\href{https://arxiv.org/abs/1904.00009}{{\ttfamily 1904.00009}}].

\bibitem{Peraro:2019svx}
T.~Peraro, \emph{{FiniteFlow: multivariate functional reconstruction using
  finite fields and dataflow graphs}},
  \href{https://doi.org/10.1007/JHEP07(2019)031}{\emph{JHEP} {\bfseries 07}
  (2019) 031} [\href{https://arxiv.org/abs/1905.08019}{{\ttfamily
  1905.08019}}].

\bibitem{Abreu:2020xvt}
S.~Abreu, J.~Dormans, F.~Febres~Cordero, H.~Ita, M.~Kraus, B.~Page et~al.,
  \emph{{Caravel: A C++ framework for the computation of multi-loop amplitudes
  with numerical unitarity}},
  \href{https://doi.org/10.1016/j.cpc.2021.108069}{\emph{Comput. Phys. Commun.}
  {\bfseries 267} (2021) 108069}
  [\href{https://arxiv.org/abs/2009.11957}{{\ttfamily 2009.11957}}].

\bibitem{Laurentis:2019bjh}
G.~Laurentis and D.~Ma\^\i{}tre, \emph{{Extracting analytical one-loop
  amplitudes from numerical evaluations}},
  \href{https://doi.org/10.1007/JHEP07(2019)123}{\emph{JHEP} {\bfseries 07}
  (2019) 123} [\href{https://arxiv.org/abs/1904.04067}{{\ttfamily
  1904.04067}}].

\bibitem{Budge:2020oyl}
L.~Budge, J.~M. Campbell, G.~De~Laurentis, R.~K. Ellis and S.~Seth, \emph{{The
  one-loop amplitudes for Higgs + 4 partons with full mass effects}},
  \href{https://doi.org/10.1007/JHEP05(2020)079}{\emph{JHEP} {\bfseries 05}
  (2020) 079} [\href{https://arxiv.org/abs/2002.04018}{{\ttfamily
  2002.04018}}].

\bibitem{Campbell:2021mlr}
J.~M. Campbell, G.~De~Laurentis, R.~K. Ellis and S.~Seth, \emph{{The pp
  \textrightarrow{} W(\textrightarrow{} l\ensuremath{\nu}) +
  \ensuremath{\gamma} process at next-to-next-to-leading order}},
  \href{https://doi.org/10.1007/JHEP07(2021)079}{\emph{JHEP} {\bfseries 07}
  (2021) 079} [\href{https://arxiv.org/abs/2105.00954}{{\ttfamily
  2105.00954}}].

\bibitem{Campbell:2022qpq}
J.~M. Campbell, G.~De~Laurentis and R.~K. Ellis, \emph{{Vector boson pair
  production at one loop: analytic results for the process $q \bar{q} \ell
  \bar\ell \ell^\prime \bar{\ell}^\prime g$}},
  \href{https://arxiv.org/abs/2203.17170}{{\ttfamily 2203.17170}}.

\bibitem{DeLaurentis:2022otd}
G.~De~Laurentis and B.~Page, \emph{{Ans\"atze for Scattering Amplitudes from
  $p$-adic Numbers and Algebraic Geometry}},
  \href{https://arxiv.org/abs/2203.04269}{{\ttfamily 2203.04269}}.

\bibitem{Pak:2011xt}
A.~Pak, \emph{{The Toolbox of modern multi-loop calculations: novel analytic
  and semi-analytic techniques}},
  \href{https://doi.org/10.1088/1742-6596/368/1/012049}{\emph{J. Phys. Conf.
  Ser.} {\bfseries 368} (2012) 012049}
  [\href{https://arxiv.org/abs/1111.0868}{{\ttfamily 1111.0868}}].

\bibitem{Raichev:2012pf}
A.~Raichev, \emph{Le{\u\i}nartas' partial fraction decomposition},
  \href{https://arxiv.org/abs/1206.4740}{{\ttfamily 1206.4740}}.

\bibitem{Bendle:2019csk}
D.~Bendle, J.~B\"ohm, W.~Decker, A.~Georgoudis, F.-J. Pfreundt, M.~Rahn et~al.,
  \emph{{Integration-by-parts reductions of Feynman integrals using Singular
  and GPI-Space}}, \href{https://doi.org/10.1007/JHEP02(2020)079}{\emph{JHEP}
  {\bfseries 02} (2020) 079}
  [\href{https://arxiv.org/abs/1908.04301}{{\ttfamily 1908.04301}}].

\bibitem{Heller:2021qkz}
M.~Heller and A.~von Manteuffel, \emph{{MultivariateApart: Generalized partial
  fractions}}, \href{https://doi.org/10.1016/j.cpc.2021.108174}{\emph{Comput.
  Phys. Commun.} {\bfseries 271} (2022) 108174}
  [\href{https://arxiv.org/abs/2101.08283}{{\ttfamily 2101.08283}}].

\bibitem{Chen:2022cgv}
X.~Chen, T.~Gehrmann, E.~W.~N. Glover, A.~Huss, P.~Monni, E.~Re et~al.,
  \emph{{Third order fiducial predictions for Drell-Yan at the LHC}},
  \href{https://arxiv.org/abs/2203.01565}{{\ttfamily 2203.01565}}.

\bibitem{Cacciari:2015jma}
M.~Cacciari, F.~A. Dreyer, A.~Karlberg, G.~P. Salam and G.~Zanderighi,
  \emph{{Fully Differential Vector-Boson-Fusion Higgs Production at
  Next-to-Next-to-Leading Order}},
  \href{https://doi.org/10.1103/PhysRevLett.115.082002}{\emph{Phys. Rev. Lett.}
  {\bfseries 115} (2015) 082002}
  [\href{https://arxiv.org/abs/1506.02660}{{\ttfamily 1506.02660}}].

\bibitem{Catani:2007vq}
S.~Catani and M.~Grazzini, \emph{{An NNLO subtraction formalism in hadron
  collisions and its application to Higgs boson production at the LHC}},
  \href{https://doi.org/10.1103/PhysRevLett.98.222002}{\emph{Phys. Rev. Lett.}
  {\bfseries 98} (2007) 222002}
  [\href{https://arxiv.org/abs/hep-ph/0703012}{{\ttfamily hep-ph/0703012}}].

\bibitem{Becher:2010tm}
T.~Becher and M.~Neubert, \emph{{Drell-Yan Production at Small $q_T$,
  Transverse Parton Distributions and the Collinear Anomaly}},
  \href{https://doi.org/10.1140/epjc/s10052-011-1665-7}{\emph{Eur. Phys. J. C}
  {\bfseries 71} (2011) 1665}
  [\href{https://arxiv.org/abs/1007.4005}{{\ttfamily 1007.4005}}].

\bibitem{Billis:2019vxg}
G.~Billis, M.~A. Ebert, J.~K.~L. Michel and F.~J. Tackmann, \emph{{A toolbox
  for $q_{T}$ and 0-jettiness subtractions at $\hbox {N}^3\hbox {LO}$}},
  \href{https://doi.org/10.1140/epjp/s13360-021-01155-y}{\emph{Eur. Phys. J.
  Plus} {\bfseries 136} (2021) 214}
  [\href{https://arxiv.org/abs/1909.00811}{{\ttfamily 1909.00811}}].

\bibitem{Stewart:2010tn}
I.~W. Stewart, F.~J. Tackmann and W.~J. Waalewijn, \emph{{N-Jettiness: An
  Inclusive Event Shape to Veto Jets}},
  \href{https://doi.org/10.1103/PhysRevLett.105.092002}{\emph{Phys. Rev. Lett.}
  {\bfseries 105} (2010) 092002}
  [\href{https://arxiv.org/abs/1004.2489}{{\ttfamily 1004.2489}}].

\bibitem{Boughezal:2015dva}
R.~Boughezal, C.~Focke, X.~Liu and F.~Petriello, \emph{{$W$-boson production in
  association with a jet at next-to-next-to-leading order in perturbative
  QCD}}, \href{https://doi.org/10.1103/PhysRevLett.115.062002}{\emph{Phys. Rev.
  Lett.} {\bfseries 115} (2015) 062002}
  [\href{https://arxiv.org/abs/1504.02131}{{\ttfamily 1504.02131}}].

\bibitem{Gaunt:2015pea}
J.~Gaunt, M.~Stahlhofen, F.~J. Tackmann and J.~R. Walsh, \emph{{N-jettiness
  Subtractions for NNLO QCD Calculations}},
  \href{https://doi.org/10.1007/JHEP09(2015)058}{\emph{JHEP} {\bfseries 09}
  (2015) 058} [\href{https://arxiv.org/abs/1505.04794}{{\ttfamily
  1505.04794}}].

\bibitem{Li:2016ctv}
Y.~Li and H.~X. Zhu, \emph{{Bootstrapping Rapidity Anomalous Dimensions for
  Transverse-Momentum Resummation}},
  \href{https://doi.org/10.1103/PhysRevLett.118.022004}{\emph{Phys. Rev. Lett.}
  {\bfseries 118} (2017) 022004}
  [\href{https://arxiv.org/abs/1604.01404}{{\ttfamily 1604.01404}}].

\bibitem{Vladimirov:2016dll}
A.~A. Vladimirov, \emph{{Correspondence between Soft and Rapidity Anomalous
  Dimensions}},
  \href{https://doi.org/10.1103/PhysRevLett.118.062001}{\emph{Phys. Rev. Lett.}
  {\bfseries 118} (2017) 062001}
  [\href{https://arxiv.org/abs/1610.05791}{{\ttfamily 1610.05791}}].

\bibitem{Luo:2020epw}
M.-x. Luo, T.-Z. Yang, H.~X. Zhu and Y.~J. Zhu, \emph{{Unpolarized quark and
  gluon TMD PDFs and FFs at N$^{3}$LO}},
  \href{https://doi.org/10.1007/JHEP06(2021)115}{\emph{JHEP} {\bfseries 06}
  (2021) 115} [\href{https://arxiv.org/abs/2012.03256}{{\ttfamily
  2012.03256}}].

\bibitem{Ebert:2020yqt}
M.~A. Ebert, B.~Mistlberger and G.~Vita, \emph{{Transverse momentum dependent
  PDFs at N$^3$LO}}, \href{https://doi.org/10.1007/JHEP09(2020)146}{\emph{JHEP}
  {\bfseries 09} (2020) 146}
  [\href{https://arxiv.org/abs/2006.05329}{{\ttfamily 2006.05329}}].

\bibitem{Caola:2022ayt}
F.~Caola, W.~Chen, C.~Duhr, X.~Liu, B.~Mistlberger, F.~Petriello et~al.,
  \emph{{The Path forward to N$^3$LO}},  in \emph{{2022 Snowmass Summer
  Study}}, 3, 2022, \href{https://arxiv.org/abs/2203.06730}{{\ttfamily
  2203.06730}}.

\bibitem{Moult:2016fqy}
I.~Moult, L.~Rothen, I.~W. Stewart, F.~J. Tackmann and H.~X. Zhu,
  \emph{{Subleading Power Corrections for N-Jettiness Subtractions}},
  \href{https://doi.org/10.1103/PhysRevD.95.074023}{\emph{Phys. Rev. D}
  {\bfseries 95} (2017) 074023}
  [\href{https://arxiv.org/abs/1612.00450}{{\ttfamily 1612.00450}}].

\bibitem{Boughezal:2018mvf}
R.~Boughezal, A.~Isgr\`o and F.~Petriello, \emph{{Next-to-leading-logarithmic
  power corrections for $N$-jettiness subtraction in color-singlet
  production}}, \href{https://doi.org/10.1103/PhysRevD.97.076006}{\emph{Phys.
  Rev. D} {\bfseries 97} (2018) 076006}
  [\href{https://arxiv.org/abs/1802.00456}{{\ttfamily 1802.00456}}].

\bibitem{Ebert:2018gsn}
M.~A. Ebert, I.~Moult, I.~W. Stewart, F.~J. Tackmann, G.~Vita and H.~X. Zhu,
  \emph{{Subleading power rapidity divergences and power corrections for
  q$_{T}$}}, \href{https://doi.org/10.1007/JHEP04(2019)123}{\emph{JHEP}
  {\bfseries 04} (2019) 123}
  [\href{https://arxiv.org/abs/1812.08189}{{\ttfamily 1812.08189}}].

\bibitem{Herzog:2018ily}
F.~Herzog, \emph{{Geometric IR subtraction for final state real radiation}},
  \href{https://doi.org/10.1007/JHEP08(2018)006}{\emph{JHEP} {\bfseries 08}
  (2018) 006} [\href{https://arxiv.org/abs/1804.07949}{{\ttfamily
  1804.07949}}].

\bibitem{Gehrmann-DeRidder:2005btv}
A.~Gehrmann-De~Ridder, T.~Gehrmann and E.~W.~N. Glover, \emph{{Antenna
  subtraction at NNLO}},
  \href{https://doi.org/10.1088/1126-6708/2005/09/056}{\emph{JHEP} {\bfseries
  09} (2005) 056} [\href{https://arxiv.org/abs/hep-ph/0505111}{{\ttfamily
  hep-ph/0505111}}].

\bibitem{Currie:2013vh}
J.~Currie, E.~W.~N. Glover and S.~Wells, \emph{{Infrared Structure at NNLO
  Using Antenna Subtraction}},
  \href{https://doi.org/10.1007/JHEP04(2013)066}{\emph{JHEP} {\bfseries 04}
  (2013) 066} [\href{https://arxiv.org/abs/1301.4693}{{\ttfamily 1301.4693}}].

\bibitem{Czakon:2010td}
M.~Czakon, \emph{{A novel subtraction scheme for double-real radiation at
  NNLO}}, \href{https://doi.org/10.1016/j.physletb.2010.08.036}{\emph{Phys.
  Lett. B} {\bfseries 693} (2010) 259}
  [\href{https://arxiv.org/abs/1005.0274}{{\ttfamily 1005.0274}}].

\bibitem{Boughezal:2011jf}
R.~Boughezal, K.~Melnikov and F.~Petriello, \emph{{A subtraction scheme for
  NNLO computations}},
  \href{https://doi.org/10.1103/PhysRevD.85.034025}{\emph{Phys. Rev. D}
  {\bfseries 85} (2012) 034025}
  [\href{https://arxiv.org/abs/1111.7041}{{\ttfamily 1111.7041}}].

\bibitem{Czakon:2014oma}
M.~Czakon and D.~Heymes, \emph{{Four-dimensional formulation of the
  sector-improved residue subtraction scheme}},
  \href{https://doi.org/10.1016/j.nuclphysb.2014.11.006}{\emph{Nucl. Phys. B}
  {\bfseries 890} (2014) 152}
  [\href{https://arxiv.org/abs/1408.2500}{{\ttfamily 1408.2500}}].

\bibitem{Currie:2017eqf}
J.~Currie, A.~Gehrmann-De~Ridder, T.~Gehrmann, E.~W.~N. Glover, A.~Huss and
  J.~Pires, \emph{{Precise predictions for dijet production at the LHC}},
  \href{https://doi.org/10.1103/PhysRevLett.119.152001}{\emph{Phys. Rev. Lett.}
  {\bfseries 119} (2017) 152001}
  [\href{https://arxiv.org/abs/1705.10271}{{\ttfamily 1705.10271}}].

\bibitem{Gehrmann-DeRidder:2019ibf}
A.~Gehrmann-De~Ridder, T.~Gehrmann, E.~W.~N. Glover, A.~Huss and J.~Pires,
  \emph{{Triple Differential Dijet Cross Section at the LHC}},
  \href{https://doi.org/10.1103/PhysRevLett.123.102001}{\emph{Phys. Rev. Lett.}
  {\bfseries 123} (2019) 102001}
  [\href{https://arxiv.org/abs/1905.09047}{{\ttfamily 1905.09047}}].

\bibitem{Czakon:2019tmo}
M.~Czakon, A.~van Hameren, A.~Mitov and R.~Poncelet, \emph{{Single-jet
  inclusive rates with exact color at $ \mathcal{O} $ ($ {\alpha}_s^4 $)}},
  \href{https://doi.org/10.1007/JHEP10(2019)262}{\emph{JHEP} {\bfseries 10}
  (2019) 262} [\href{https://arxiv.org/abs/1907.12911}{{\ttfamily
  1907.12911}}].

\bibitem{Czakon:2021mjy}
M.~Czakon, A.~Mitov and R.~Poncelet, \emph{{Next-to-Next-to-Leading Order Study
  of Three-Jet Production at the LHC}},
  \href{https://doi.org/10.1103/PhysRevLett.127.152001}{\emph{Phys. Rev. Lett.}
  {\bfseries 127} (2021) 152001}
  [\href{https://arxiv.org/abs/2106.05331}{{\ttfamily 2106.05331}}].

\bibitem{Chen:2022ktf}
X.~Chen, T.~Gehrmann, N.~Glover, A.~Huss and M.~Marcoli, \emph{{Automation of
  antenna subtraction in colour space: gluonic processes}},
  \href{https://arxiv.org/abs/2203.13531}{{\ttfamily 2203.13531}}.

\bibitem{Caola:2017dug}
F.~Caola, K.~Melnikov and R.~R\"ontsch, \emph{{Nested soft-collinear
  subtractions in NNLO QCD computations}},
  \href{https://doi.org/10.1140/epjc/s10052-017-4774-0}{\emph{Eur. Phys. J. C}
  {\bfseries 77} (2017) 248}
  [\href{https://arxiv.org/abs/1702.01352}{{\ttfamily 1702.01352}}].

\bibitem{Asteriadis:2019dte}
K.~Asteriadis, F.~Caola, K.~Melnikov and R.~R\"ontsch, \emph{{Analytic results
  for deep-inelastic scattering at NNLO QCD with the nested soft-collinear
  subtraction scheme}},
  \href{https://doi.org/10.1140/epjc/s10052-019-7567-9}{\emph{Eur. Phys. J. C}
  {\bfseries 80} (2020) 8} [\href{https://arxiv.org/abs/1910.13761}{{\ttfamily
  1910.13761}}].

\bibitem{Asteriadis:2021gpd}
K.~Asteriadis, F.~Caola, K.~Melnikov and R.~R\"ontsch, \emph{{NNLO QCD
  corrections to weak boson fusion Higgs boson production in the H
  \textrightarrow{} b$ \overline{b} $ and H \textrightarrow{} WW$^{*}$
  \textrightarrow{} 4l decay channels}},
  \href{https://doi.org/10.1007/JHEP02(2022)046}{\emph{JHEP} {\bfseries 02}
  (2022) 046} [\href{https://arxiv.org/abs/2110.02818}{{\ttfamily
  2110.02818}}].

\bibitem{Buccioni:2022kgy}
F.~Buccioni, F.~Caola, H.~A. Chawdhry, F.~Devoto, M.~Heller, A.~von Manteuffel
  et~al., \emph{{Mixed QCD-electroweak corrections to dilepton production at
  the LHC in the high invariant mass region}},
  \href{https://arxiv.org/abs/2203.11237}{{\ttfamily 2203.11237}}.

\bibitem{Somogyi:2005xz}
G.~Somogyi, Z.~Trocsanyi and V.~Del~Duca, \emph{{Matching of singly- and
  doubly-unresolved limits of tree-level QCD squared matrix elements}},
  \href{https://doi.org/10.1088/1126-6708/2005/06/024}{\emph{JHEP} {\bfseries
  06} (2005) 024} [\href{https://arxiv.org/abs/hep-ph/0502226}{{\ttfamily
  hep-ph/0502226}}].

\bibitem{DelDuca:2016ily}
V.~Del~Duca, C.~Duhr, A.~Kardos, G.~Somogyi, Z.~Sz\H{o}r, Z.~Tr\'ocs\'anyi
  et~al., \emph{{Jet production in the CoLoRFulNNLO method: event shapes in
  electron-positron collisions}},
  \href{https://doi.org/10.1103/PhysRevD.94.074019}{\emph{Phys. Rev. D}
  {\bfseries 94} (2016) 074019}
  [\href{https://arxiv.org/abs/1606.03453}{{\ttfamily 1606.03453}}].

\bibitem{DelDuca:2015zqa}
V.~Del~Duca, C.~Duhr, G.~Somogyi, F.~Tramontano and Z.~Tr\'ocs\'anyi,
  \emph{{Higgs boson decay into b-quarks at NNLO accuracy}},
  \href{https://doi.org/10.1007/JHEP04(2015)036}{\emph{JHEP} {\bfseries 04}
  (2015) 036} [\href{https://arxiv.org/abs/1501.07226}{{\ttfamily
  1501.07226}}].

\bibitem{Magnea:2018hab}
L.~Magnea, E.~Maina, G.~Pelliccioli, C.~Signorile-Signorile, P.~Torrielli and
  S.~Uccirati, \emph{{Local analytic sector subtraction at NNLO}},
  \href{https://doi.org/10.1007/JHEP12(2018)107}{\emph{JHEP} {\bfseries 12}
  (2018) 107} [\href{https://arxiv.org/abs/1806.09570}{{\ttfamily
  1806.09570}}].

\bibitem{Heinrich:2020ybq}
G.~Heinrich, \emph{{Collider Physics at the Precision Frontier}},
  \href{https://doi.org/10.1016/j.physrep.2021.03.006}{\emph{Phys. Rept.}
  {\bfseries 922} (2021) 1} [\href{https://arxiv.org/abs/2009.00516}{{\ttfamily
  2009.00516}}].

\bibitem{Chen:2021isd}
X.~Chen, T.~Gehrmann, E.~W.~N. Glover, A.~Huss, B.~Mistlberger and A.~Pelloni,
  \emph{{Fully Differential Higgs Boson Production to Third Order in QCD}},
  \href{https://doi.org/10.1103/PhysRevLett.127.072002}{\emph{Phys. Rev. Lett.}
  {\bfseries 127} (2021) 072002}
  [\href{https://arxiv.org/abs/2102.07607}{{\ttfamily 2102.07607}}].

\bibitem{Billis:2021ecs}
G.~Billis, B.~Dehnadi, M.~A. Ebert, J.~K.~L. Michel and F.~J. Tackmann,
  \emph{{Higgs pT Spectrum and Total Cross Section with Fiducial Cuts at Third
  Resummed and Fixed Order in QCD}},
  \href{https://doi.org/10.1103/PhysRevLett.127.072001}{\emph{Phys. Rev. Lett.}
  {\bfseries 127} (2021) 072001}
  [\href{https://arxiv.org/abs/2102.08039}{{\ttfamily 2102.08039}}].

\bibitem{Chawdhry:2019bji}
H.~A. Chawdhry, M.~L. Czakon, A.~Mitov and R.~Poncelet, \emph{{NNLO QCD
  corrections to three-photon production at the LHC}},
  \href{https://doi.org/10.1007/JHEP02(2020)057}{\emph{JHEP} {\bfseries 02}
  (2020) 057} [\href{https://arxiv.org/abs/1911.00479}{{\ttfamily
  1911.00479}}].

\bibitem{Kallweit:2020gcp}
S.~Kallweit, V.~Sotnikov and M.~Wiesemann, \emph{{Triphoton production at
  hadron colliders in NNLO QCD}},
  \href{https://doi.org/10.1016/j.physletb.2020.136013}{\emph{Phys. Lett. B}
  {\bfseries 812} (2021) 136013}
  [\href{https://arxiv.org/abs/2010.04681}{{\ttfamily 2010.04681}}].

\bibitem{Chawdhry:2021hkp}
H.~A. Chawdhry, M.~Czakon, A.~Mitov and R.~Poncelet, \emph{{NNLO QCD
  corrections to diphoton production with an additional jet at the LHC}},
  \href{https://doi.org/10.1007/JHEP09(2021)093}{\emph{JHEP} {\bfseries 09}
  (2021) 093} [\href{https://arxiv.org/abs/2105.06940}{{\ttfamily
  2105.06940}}].

\bibitem{Badger:2021ohm}
S.~Badger, T.~Gehrmann, M.~Marcoli and R.~Moodie, \emph{{Next-to-leading order
  QCD corrections to diphoton-plus-jet production through gluon fusion at the
  LHC}}, \href{https://doi.org/10.1016/j.physletb.2021.136802}{\emph{Phys.
  Lett. B} {\bfseries 824} (2022) 136802}
  [\href{https://arxiv.org/abs/2109.12003}{{\ttfamily 2109.12003}}].

\bibitem{Badger:2021imn}
S.~Badger, C.~Br\o{}nnum-Hansen, D.~Chicherin, T.~Gehrmann, H.~B. Hartanto,
  J.~Henn et~al., \emph{{Virtual QCD corrections to gluon-initiated diphoton
  plus jet production at hadron colliders}},
  \href{https://doi.org/10.1007/JHEP11(2021)083}{\emph{JHEP} {\bfseries 11}
  (2021) 083} [\href{https://arxiv.org/abs/2106.08664}{{\ttfamily
  2106.08664}}].

\bibitem{Pellen:2021vpi}
M.~Pellen, R.~Poncelet and A.~Popescu, \emph{{Polarised W+j production at the
  LHC: a study at NNLO QCD accuracy}},
  \href{https://doi.org/10.1007/JHEP02(2022)160}{\emph{JHEP} {\bfseries 02}
  (2022) 160} [\href{https://arxiv.org/abs/2109.14336}{{\ttfamily
  2109.14336}}].

\bibitem{Czakon:2020qbd}
M.~Czakon, A.~Mitov and R.~Poncelet, \emph{{NNLO QCD corrections to leptonic
  observables in top-quark pair production and decay}},
  \href{https://doi.org/10.1007/JHEP05(2021)212}{\emph{JHEP} {\bfseries 05}
  (2021) 212} [\href{https://arxiv.org/abs/2008.11133}{{\ttfamily
  2008.11133}}].

\bibitem{Czakon:2021ohs}
M.~L. Czakon, T.~Generet, A.~Mitov and R.~Poncelet, \emph{{B-hadron production
  in NNLO QCD: application to LHC t$ \overline{t} $ events with leptonic
  decays}}, \href{https://doi.org/10.1007/JHEP10(2021)216}{\emph{JHEP}
  {\bfseries 10} (2021) 216}
  [\href{https://arxiv.org/abs/2102.08267}{{\ttfamily 2102.08267}}].

\bibitem{Alekhin:2021xcu}
S.~Alekhin, A.~Kardos, S.~Moch and Z.~Tr\'ocs\'anyi, \emph{{Precision studies
  for Drell\textendash{}Yan processes at NNLO}},
  \href{https://doi.org/10.1140/epjc/s10052-021-09361-9}{\emph{Eur. Phys. J. C}
  {\bfseries 81} (2021) 573}
  [\href{https://arxiv.org/abs/2104.02400}{{\ttfamily 2104.02400}}].

\bibitem{Behring:2020cqi}
A.~Behring, F.~Buccioni, F.~Caola, M.~Delto, M.~Jaquier, K.~Melnikov et~al.,
  \emph{{Mixed QCD-electroweak corrections to $W$-boson production in hadron
  collisions}}, \href{https://doi.org/10.1103/PhysRevD.103.013008}{\emph{Phys.
  Rev. D} {\bfseries 103} (2021) 013008}
  [\href{https://arxiv.org/abs/2009.10386}{{\ttfamily 2009.10386}}].

\bibitem{Bevilacqua:2021cit}
G.~Bevilacqua, H.-Y. Bi, H.~B. Hartanto, M.~Kraus, M.~Lupattelli and M.~Worek,
  \emph{{$ t\bar{t}b\bar{b} $ at the LHC: on the size of corrections and b-jet
  definitions}}, \href{https://doi.org/10.1007/JHEP08(2021)008}{\emph{JHEP}
  {\bfseries 08} (2021) 008}
  [\href{https://arxiv.org/abs/2105.08404}{{\ttfamily 2105.08404}}].

\bibitem{Bevilacqua:2022twl}
G.~Bevilacqua, H.-Y. Bi, H.~B. Hartanto, M.~Kraus, M.~Lupattelli and M.~Worek,
  \emph{{$t\bar{t}b\bar{b}$ at the LHC: On the size of off-shell effects and
  prompt $b$-jet identification}},
  \href{https://arxiv.org/abs/2202.11186}{{\ttfamily 2202.11186}}.

\bibitem{Denner:2020orv}
A.~Denner, J.-N. Lang and M.~Pellen, \emph{{Full NLO QCD corrections to
  off-shell tt\textasciimacron{}bb\textasciimacron{} production}},
  \href{https://doi.org/10.1103/PhysRevD.104.056018}{\emph{Phys. Rev. D}
  {\bfseries 104} (2021) 056018}
  [\href{https://arxiv.org/abs/2008.00918}{{\ttfamily 2008.00918}}].

\bibitem{Campbell:2022qmc}
J.~M. Campbell et~al., \emph{{Event Generators for High-Energy Physics
  Experiments}},  in \emph{{2022 Snowmass Summer Study}}, 3, 2022,
  \href{https://arxiv.org/abs/2203.11110}{{\ttfamily 2203.11110}}.

\bibitem{Mazzitelli:2021mmm}
J.~Mazzitelli, P.~F. Monni, P.~Nason, E.~Re, M.~Wiesemann and G.~Zanderighi,
  \emph{{Top-pair production at the LHC with MiNNLO$_{\rm PS}$}},
  \href{https://arxiv.org/abs/2112.12135}{{\ttfamily 2112.12135}}.

\bibitem{Lombardi:2020wju}
D.~Lombardi, M.~Wiesemann and G.~Zanderighi, \emph{{Advancing M\i{}NNLO$_{PS}$
  to diboson processes: Z\ensuremath{\gamma} production at NNLO+PS}},
  \href{https://doi.org/10.1007/JHEP06(2021)095}{\emph{JHEP} {\bfseries 06}
  (2021) 095} [\href{https://arxiv.org/abs/2010.10478}{{\ttfamily
  2010.10478}}].

\bibitem{Cridge:2021hfr}
T.~Cridge, M.~A. Lim and R.~Nagar, \emph{{W\ensuremath{\gamma} production at
  NNLO+PS accuracy in Geneva}},
  \href{https://doi.org/10.1016/j.physletb.2022.136918}{\emph{Phys. Lett. B}
  {\bfseries 826} (2022) 136918}
  [\href{https://arxiv.org/abs/2105.13214}{{\ttfamily 2105.13214}}].

\bibitem{Alioli:2021egp}
S.~Alioli, A.~Broggio, A.~Gavardi, S.~Kallweit, M.~A. Lim, R.~Nagar et~al.,
  \emph{{Next-to-next-to-leading order event generation for $Z$ boson pair
  production matched to parton shower}},
  \href{https://doi.org/10.1016/j.physletb.2021.136380}{\emph{Phys. Lett. B}
  {\bfseries 818} (2021) 136380}
  [\href{https://arxiv.org/abs/2103.01214}{{\ttfamily 2103.01214}}].

\bibitem{Butter:2022rso}
A.~Butter et~al., \emph{{Machine Learning and LHC Event Generation}},  in
  \emph{{2022 Snowmass Summer Study}}, 3, 2022,
  \href{https://arxiv.org/abs/2203.07460}{{\ttfamily 2203.07460}}.

\end{thebibliography}\endgroup
	
\end{document}